\documentclass{article}

\title{Numerical simulations of elastic capsules with nucleus in shear flow}
\author{Arash Alizad Banaei, Jean-Christophe Loiseau, Iman Lashgari and Luca Brandt \\
Linn\'e FLOW Centre and SeRC (Swedish e-science Research Centre), \\KTH Mechanics, SE 10044, Stockholm, Sweden}
\date{}

\usepackage{fullpage}
\usepackage{graphicx}
  \usepackage{epstopdf}
\usepackage[]{amsmath}
\usepackage[]{xcolor}
\usepackage[]{subfigure}
\graphicspath{ {figures/} }

\begin{document}

\maketitle

\begin{abstract}

The shear-induced deformation of a capsule with a stiff nucleus, a model of eukaryotic cells, is studied numerically. The membrane of the cell and of its nucleus are modelled as a thin and impermeable elastic material obeying a Neo-Hookean constitutive law. The membranes are discretised by a Lagrangian mesh and their governing equations are solved in spectral space using spherical harmonics, while the fluid equations are solved on a staggered grid using a second-order finite differences scheme. The fluid-structure coupling is obtained using an immersed boundary method. The numerical approach is presented and validated for the case of a single capsule in a shear flow. The variations induced by the presence of the nucleus on the cell deformation are investigated when varying the viscosity ratio between the inner and outer fluids, the membrane elasticity and its bending stiffness. The deformation of the eukaryotic cell is smaller than that of the prokaryotic one. The reduction in deformation increases for larger values of the capillary number. The eukaryotic cell remains thicker in its middle part compared to the prokaryotic one, thus making it less flexible to pass through narrow capillaries. For a viscosity ratio of 5, the deformation of the cell is smaller than in the case of uniform viscosity. In addition, for non-zero bending stiffness of the membrane, the deformation decreases and the shape is closer to an ellipsoid. Finally, we compare the results obtained modeling the nucleus as an inner stiffer membrane with those obtained using a rigid particle.

\end{abstract}

\section{Introduction}

The deformation of a cell in shear flows is one of the fundamental mechanical problems in cell biology. A living cell is subjected to mechanical forces of various magnitude, direction and distribution throughout its life. The cell response to those forces reflects its biological function \cite{lim2006mechanical}. As an example, red blood cells (RBC) have a diameter of about $7 \mu m$ in the underformed state. Their ability to deform quite significantly allow them to pass through narrow capillaries having a diameter of $3 \mu m$ \cite{wu2013simulation}. This high deformability enables them to reach various parts of the human body and to distribute oxygen and nutrient to cells. RBC can however be affected by \textit{protozoan Plamodium falciparum}, the parasites that cause malaria. These parasites change the red blood cell chemical and structural composition \cite{bannister2003ins,cooke2001malaria}, thus inducing a stiffening of the cell membrane \cite{zhang2015multiple,guo2016deformability}. Such variations in the mechanical properties of the cells affects the blood rheology which may help to diagnose diseases.

Many early experimental studies have addressed the interaction between tiny deformable particles and an external flow. Several interesting types of motion have been discovered such as {\it tumbling} and {\it tank-treading} in shear flow \cite{goldsmith1972flow,fischer1977tank}, the \textit{zipper} flow pattern \cite{gaehtgens1979motion} or parachute cell shapes \cite{skalak1969deformation}. More recent studies focused on cells that exhibit very large deformations at high shear rates, which can cause breaking \cite{chang1993experimental}, just to mention few examples. Most of these studies are of experimental nature. Such investigations can however be quite expensive since they require dedicated facilities not easy to fabricate. In addition, experimentally measuring the exact deformation and stresses can be rather complicated. Developing robust and reliable numerical platforms is thus of increasing importance in order to perform high-fidelity simulations beside laboratory experiments.

 Many cells, including red blood cells, can be modeled as capsules. Capsules consist of a droplet enclosed by a thin membrane: the membrane area can vary while the enclosed volume is constant. Nowadays, several numerical studies on the deformation of a capsule in shear flow have been reported in literature. At certain shear rates, the capsule reaches a steady shape while its membrane exhibits a rotation known as tank-treading motion \cite{huang2012three}. This tank-treading motion disappears when the viscosity or shear rate of the external fluid becomes low enough and instead a flipping or tumbling motion similar to that of a rigid body appears \cite{schmid1969fluid,fischer1978red}. Membranes can also undergo buckling or folding for high elastic moduli or at low and high shear rates in absence of bending rigidity \cite{walter2001shear,huang2012three}. A solution to this problem is proposed by introducing a stress on undeformed membrane, the so-called pre-stressed capsule \cite{lac2005deformation}. As regards the motion of non-spherical capsules in shear flow, different types of motion occur when changing the fluid viscosity, the membrane elasticity, the geometry of the problem or the applied shear rate. In  \cite{skotheim2007red}, a phase diagram is presented for biconcave shaped capsule in which the transition from tank-treading to tumbling motion is identified when decreasing the shear rate. 

 For eukaryotic cells, the overall mechanical properties of a cell are not only determined by its membrane but also by other cell organelle such as the cell nucleus \cite{rodriguez2013review}. Typically, the nucleus is stiffer than the surrounding cytoplasm which results in lower deformation when subject to the external stimuli \cite{caille2002contribution,guilak2000mechanical}. To model and predict the cell behavior, the mechanical properties of the nucleus need to be quantified. To this end, both experimental tests and numerical simulations have been carried out in \cite{caille2002contribution}. The elastic modulus of the nucleus in round and spread cells was found to be around $5000 N/m^2$, roughly ten times larger than for the cytoplasm. As further example, the nucleus of bovine cells is nine times stiffer than the cytoplasm \cite{maniotis1997demonstration}, yet small deformations of the nucleus may occur when a cell is subjected to flow \cite{galbraith1998shear}. Though it can exhibit large deformation on a substrate when highly compressed, stretched or flattened  \cite{guilak1995compression,caille1998assessment,ingber1990fibronectin}, the nucleus may be assumed as a rigid particle for an intermediate range of the applied forces (external shear).

 The objective of present research is to quantify the deformation of a cell with nucleus under simple shear flow at low but finite Reynolds numbers (fluid inertia) using numerical simulations. For the present model, the cell is made of two thin (two-dimensional) elastic membranes as originally proposed by \cite{kan1999effects}: the outer membrane separating the cell from the ambient fluid and an inner membrane acting as the boundary of the nucleus. The neo-Hookean hyperelastic model is chosen for the strain energy of the membranes, while the inner and outer fluids are assumed to be Newtonian and can have different viscosities. The viscosity inside nucleus is assumed to be the same as the surrounding cytoplasm. As the deformation of the nucleus is assumed to be negligible, this is numerically treated either with a second inner stiffer membrane or, alternatively, as a rigid particle using a different numerical approach \cite{lashgari2014laminar,ardekani2016numerical}. The presence of a nucleus and the effect of viscosity ratio and bending stiffness on the cell deformation are investigated to document the potential of the numerical method here developed. The results are expected to provide an interesting comparison between mechanical properties of cells with and without nucleus.
 
 This manuscript is organised as follows. In \textsection \ref{sec: problem statement}, the geometry of the problem and the governing equations for the membranes and flow dynamics are presented. Section \textsection \ref{sec: numerical methods} provides a brief introduction to the numerical methods used in order to simulate the problem. Validations of the present implementations are presented in \textsection \ref{sec: validations} while section \textsection \ref{sec: results} reports the main results of the present study. Finally, conclusions and perspectives are given in section \textsection \ref{sec: conclusion}.

%

%The immersed boundary method is used here to couple the fluid and solid motion. In the IBM, a Lagrangian mesh is used to define the interface and compute the fluid/structure interactions. The forces from the solid or deformable body are added to the momentum equations for the fluid phase which are solved on a fixed Eulerian grid, thus removing the need to remeshing following the motion of the object. The  shape  of the  membrane,  its  deformation  and  internal  stresses are  represented  by  means  of  spherical  harmonics  to have an accurate computation of the high-order derivatives of the membrane geometry. As the spacing and number of the Lagrangian points defining the elastic membrane should be of the same order of the underlying Eulerian grid, the computation of the membrane deformation becomes very expensive when the underlying mesh is very fine. To improve efficiency, especially in the case of small deformations of stiff membranes, we introduce here two distinct sets of Lagrangian points. Using the same spectral interpolation, based on the spherical harmonics, we adopt a coarser mesh to compute the membrane deformation and internal stresses, and finer one to obtain the forces exchanged with the fluid.

\section{Problem statement}
\label{sec: problem statement}

\begin{figure}
\centering
\includegraphics[width=10cm, height=6cm]{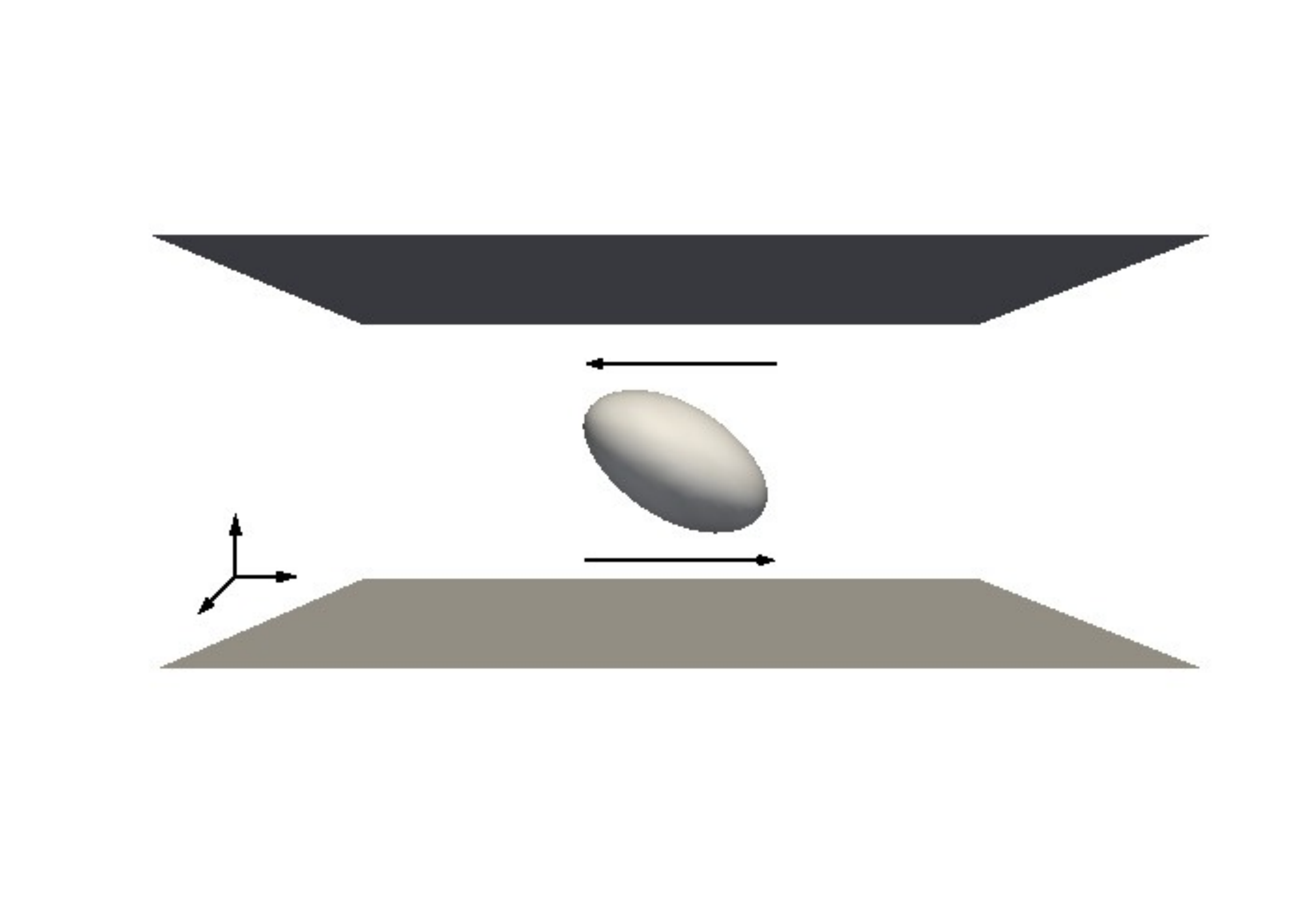}
\put(-250,54){\footnotesize $x_1$}
\put(-220,62){\footnotesize $x_2$}
\put(-236,74){\footnotesize $x_3$}
\caption{Schematic of the configuration and reference frame adopted to study a cell deformation in shear flow.}
\label{geo}
\end{figure}

    Figure \ref{geo} depicts the flow configuration considered and the coordinate system adopted. An initially spherical cell located at the geometrical center of a rectangular computation box is considered. The upper and lower walls of the domain move at opposite velocities in the streamwise direction while periodicity is assumed in the other two directions.
  As boundary conditions we therefore impose no-slip at the walls and periodicity in the steamwise and spanwise directions .

    \subsection{Navier-Stokes equations}
    
    The dynamics of the incompressible flow of a Newtonian fluid are governed by the Navier-Stokes equations,
    \begin{equation}
        \begin{aligned}
        & \nabla \cdot \textbf{u}=0, \\
        & \frac{\partial{\textbf{u}}}{\partial{t}}+\textbf{u}\cdot\nabla\textbf{u}=-\nabla P+\frac{1}{Re}\nabla\cdot\left[\mu^\ast(\nabla\textbf{u}+\nabla\textbf{u}^T)\right]  +\textbf{f},
        \end{aligned}
        \label{eq4}
    \end{equation}
        \noindent where $\textbf{u}=(u, v, w)^T$ is the velocity vector, $P$ the hydrodynamic pressure, and ${\bf f}$ indicates the fluid-solid interaction force. The Reynolds number $Re$ is defined as
    \begin{equation}
    Re=\frac{\rho \dot{\gamma}R^2}{\mu_o}.
    \end{equation}
    In the expression above, $\dot{\gamma}$ is shear rate, R the reference cell radius, $\rho$ the reference density (assumed here to be the same for the fluid inside and outside the cell) and $\mu_o$ the reference viscosity. In the present work, the reference viscosity is set to be the viscosity of the fluid outside the cell and the ratio between inner and outer viscosity is defined as $\lambda=\frac{\mu_i}{\mu_o}$, with $\mu_i$ the inner viscosity. In the expression above $\mu^\ast=\frac{\mu(\bf{x})}{\mu_o}$ indicates the ratio of viscosity at each point to the reference viscosity.

    \subsection{Membrane dynamics}
    
    Cells are surrounded by a deformable membrane known as the plasma or cytoplasmic membrane. Along with a number of biological functions, its main purpose is to separate the interior of each cell from the external environment. It can moreover deform quite significantly, as in the case of red blood cells traveling through capillary vessels. In the present numerical study, such membrane is modelled using a hyper-elastic model.

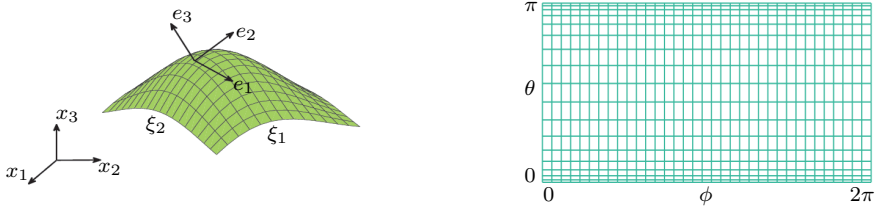
\begin{figure}
    \centering
    
    \subfigure{
    \includegraphics[width=5cm, height=3cm]{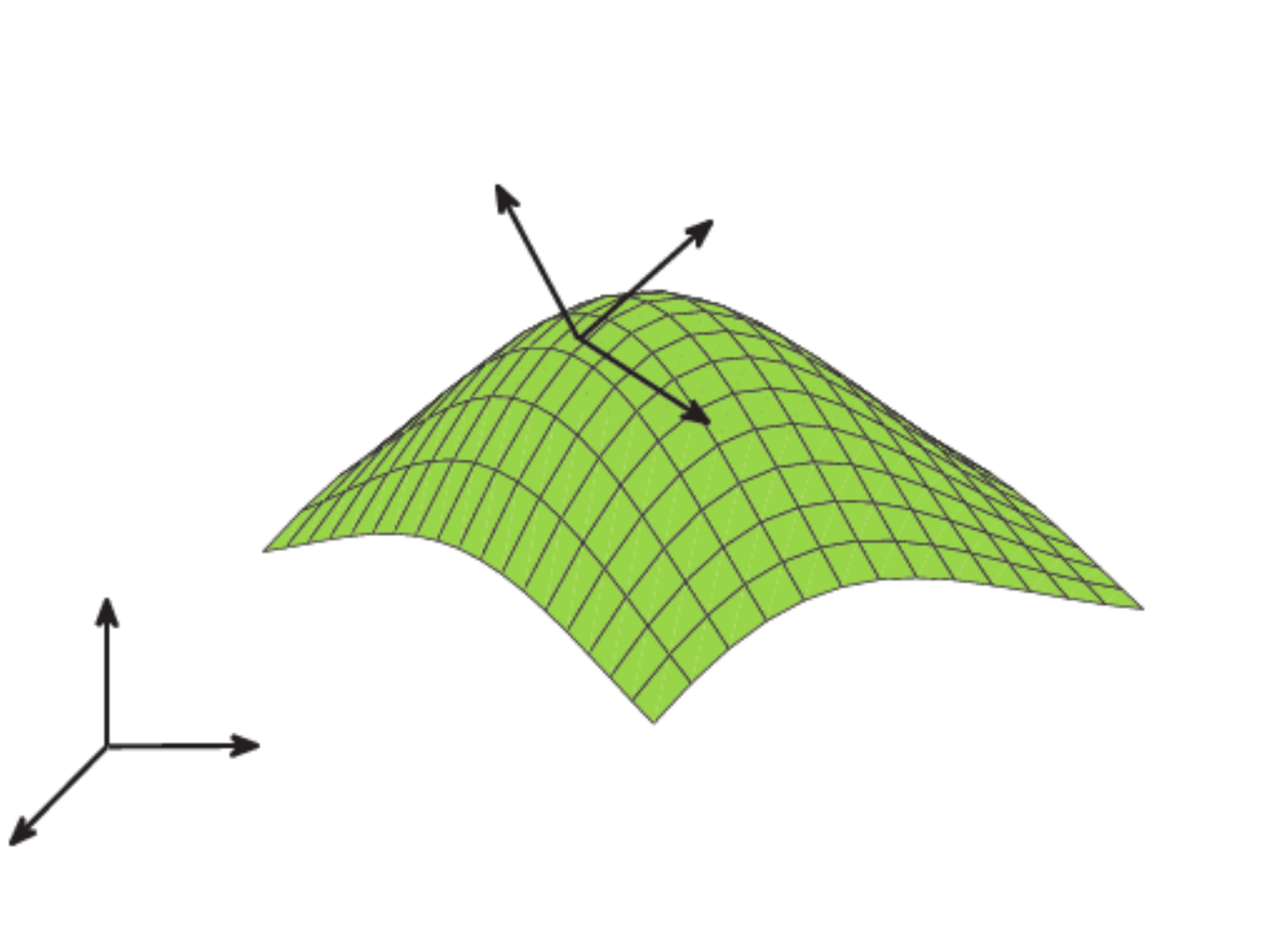}
}
\put(-153,10){\footnotesize $x_1$}
\put(-118,12){\footnotesize $x_2$}
\put(-134,32){\footnotesize $x_3$}
\put(-67,43){\footnotesize $e_1$}
\put(-66,63){\footnotesize $e_2$}
\put(-90,70){\footnotesize $e_3$}
\put(-54,24){\footnotesize $\xi_1$}
\put(-100,28){\footnotesize $\xi_2$}
    \subfigure{
    \includegraphics[width=8cm, height=3cm]{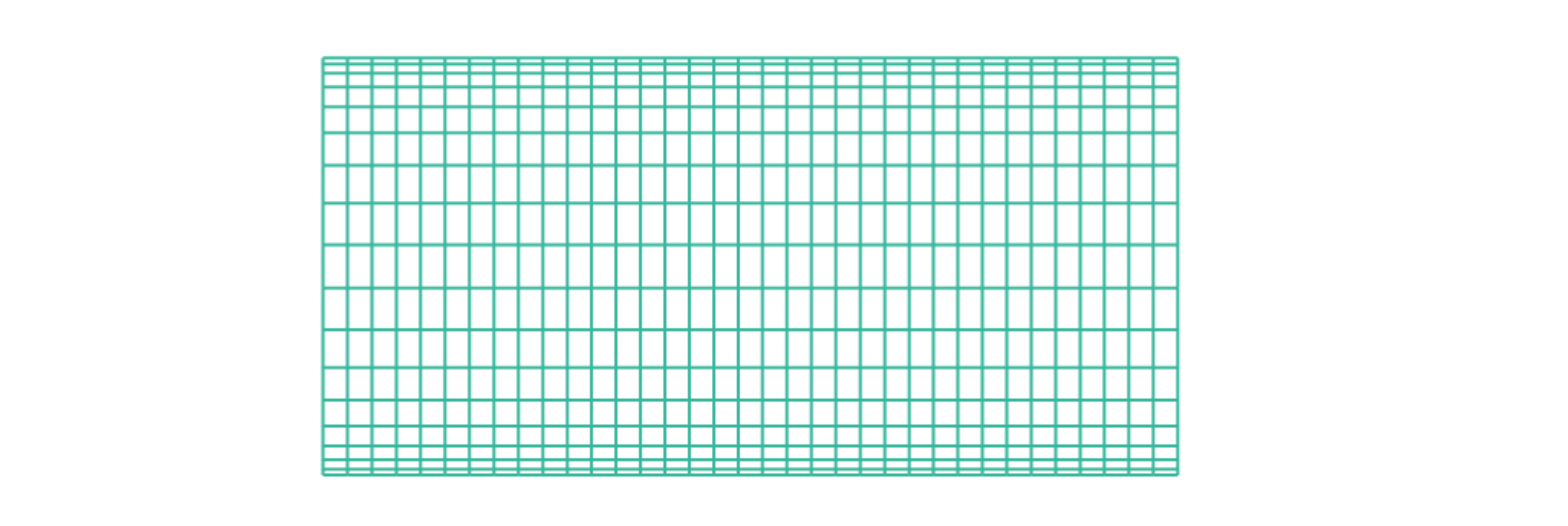}
}
\put(-125,1){\footnotesize $\phi$}
\put(-191,41){\footnotesize $\theta$}
\put(-191,8){\footnotesize $0$}
\put(-191,73){\footnotesize $\pi$}
\put(-184,1){\footnotesize $0$}
\put(-68,1){\footnotesize $2\pi$}
    \caption {Schematic of curvlinear coordinates and corresponding mesh on the cell surface}
        \label{geog}
\end{figure}

    A point on the surface of the cell is expressed by using  the curvilinear coordinates $(\xi^1,\xi^2)$.  To define the cell, two different coordinate bases are used, see figure \ref{geog}.
    The first is a fixed cartesian base, $\left({\bf e_1,e_2,e_3}\right)$ corresponding to position  $\textbf{x}\left(\xi^1,\xi^2\right)$.  The second coordinate is a local covariant base $({\bf a}_1,{\bf a}_2,{\bf a}_3)$ which follows the local deformation of the membrane. The unit vectors of the local base are
    \begin{equation} \label{a1}
        \textbf{a}_1=\frac{\partial{\textbf{x}}}{\partial{\theta}},\     \textbf{a}_2=\frac{\partial{\textbf{x}}}{\partial{\phi}}, \ \textbf{a}_3=\frac{\textbf{a}_1\times \textbf{a}_2}{\left|{\textbf{a}_1\times \textbf{a}_2}\right|},     
    \end{equation}
   where $\theta$ and $\phi$ are latitudinal and longitudinal angles on the cell surface. The co-variant and contra-variant metric tensors are defined as
    \begin{equation}
        \textbf{a}_{\alpha\beta}=\textbf{a}_\alpha\cdot\textbf{a}_\beta, \ \textbf{a}^{\alpha\beta}=\textbf{a}^\alpha\cdot\textbf{a}^\beta,     
    \end{equation}
where $\alpha,\beta=1,2$. The basis vectors and metric tensors in the undeformed (reference) state are hereafter denoted by capital letters $(\textbf{A}^\alpha,\ \textbf{A}^{\alpha\beta})$. The invariants of the transformation $I_1$ and $I_2$ are defined as
    \begin{equation}
    I_1=\textbf{\textbf{A}}^{\alpha\beta}\textbf{a}_{\alpha\beta}-2,\ I_2=\left|\textbf{A}^{\alpha\beta}\right|\left|\textbf{a}_{\alpha\beta}\right|-1.
    \end{equation}
Equivalently, they can also be determined from the principal stretching ratios $\lambda_1$ and $\lambda_2$ as
    \begin{equation}
        \begin{aligned}
        & I_1 = \lambda_1^2 +\lambda_2^2 - 2,\\
        & I_2 = \lambda_1^2 \lambda_2^2 - 1 = J_s^2 - 1.
        \end{aligned}
    \end{equation}

    The ratio of the deformed to the undeformed surface area is defined  by the Jacobian $J_s=\lambda_1\lambda_2$. The two dimensional Cauchy stress tensor, ${\bf T}$, is computed from the strain energy function per unit area $W_s(I_1,I_2)$ of the undeformed membrane as
     \begin{equation} \label{eq1}
     \textbf{T}=\frac{\textbf{1}}{\textbf{J}_s}\textbf{F}\cdot\frac{\partial{\textbf{W}_s}}{\partial{\textbf{e}}}\ \cdot\textbf{F}^T 
     \end{equation}
    \noindent where $\textbf{F}= \textbf{a}_\alpha \bigotimes \textbf{A}^\alpha$ and $\textbf{e}=(\textbf{F}^T\cdot\ \textbf{F}-\textbf{I})/2$ is the Green-Lagrange strain tensor. Equation (\ref{eq1}) can be further expressed component-wise as
    \begin{equation}
    T^{\alpha\beta}=\frac{2}{J_s}\frac{\partial{W_s}}{\partial{I_1}}\textbf{A}^{\alpha\beta}+2J_s\frac{\partial{W_s}}{\partial{I_2}}\textbf{a}^{\alpha\beta}.
    \end{equation}

    \noindent In the rest of this work, the strain energy function $W_s$ is modeled using the neo-Hookean (NH) law \cite{pozrikidis2010computational, pranay2010pair, zhu2015motion}. Using this model, the strain energy function is expressed as
    \begin{equation} \label{a3}
    W^{NH}_s= \frac{1}{2 We}\left(I_1-1+\frac{1}{I_2+1}\right),
    \end{equation}
    \noindent where $We=\frac{\rho R^3 \dot{\gamma}^2}{G_s}$ is the Weber number (or non-dimensional surface shear modulus). The local equilibrium relates the tensor $\textbf{T}$ to the external elastic load $\textbf{q}$ according to
    \begin{equation} \label{eq2}
    \nabla_s\cdot\textbf{T}+\textbf{q}=0,
    \end{equation}
    \noindent with $\nabla_s\cdot$ the surface divergence operator. In curvilinear coordinates, the load vector can be written as ${\bf q}={q}^\beta \textbf{a}_\beta+ {q}^n \textbf{n}$. The force balance in equation (\ref{eq2}) is further decomposed into tangential and normal components,
    \begin{equation}
        \begin{aligned}
        & \frac{\partial{T}^{\alpha\beta}}{\partial{\xi}^\alpha}+\Gamma^\alpha_{\alpha\lambda}T^{\lambda\beta}+\Gamma^\beta_{\alpha\lambda}T^{\alpha\lambda}+q^\beta=0, \ \ \  \beta=1,2 \\
        & T^{\alpha\beta}b_{\alpha\beta}+q^n=0
        \end{aligned}
    \end{equation}
    where $\Gamma_{\alpha\lambda}^\alpha$ and $\Gamma^\beta_{\alpha\lambda}$ are the Christoffel symbols.
    
In some cases, due to noninfinitesimal membrane thickness or a preferred configuration of an interfacial molecular network, bending moments accompanied by transverse shear tensions play an important role on cell deformation \cite{pozrikidis2001effect}. Bending stiffness is incorporated into the model using a linear isotropic model for the bending moment \cite{pozrikidis2010computational,pozrikidis2001effect}: 
    \begin{equation} \label{a4}
        \textbf{M}^{\alpha}_{\beta}=-B\left(b^\alpha_\beta-B^\alpha_\beta\right),
    \end{equation}
    \noindent where $B=\frac{G_b}{\rho R^5 \dot{\gamma}^2}$ is the non dimensional bending modulus, and $b^\alpha_\beta$ is the Gaussian curvature ($B^\alpha_\beta$ corresponds to that of the reference configuration). According to the local torque balance, including bending moments on the membrane, we obtain the transverse shear vector $\textbf{Q}$ and in-plane stress tensor $\textbf{T}$,
    \begin{equation}
        \begin{aligned}
        & {M}^{\alpha\beta}_{|\alpha}-{Q}^\beta=0, \\
        & {\boldmath{\varepsilon}}_{\alpha\beta}\left(T^{\alpha\beta}-b^\alpha_\gamma M^{\gamma\beta}\right)=0,
        \end{aligned}
        \label{eq3}
    \end{equation}
    \noindent where $^,\ _{|\alpha}\ ^,$ represents the covariant derivative and $\varepsilon$ is the two-dimensional Levi-Civita tensor. The left hand side of equation (\ref{eq3}) identifies the antisymmetric part of the in-plane stress tensor, which is always zero as proved in \cite{zhao2010spectral}. Including the transverse shear stress Q, the local stress equilibrium, including bending finally gives
    \begin{equation} \label{a2}
        \begin{aligned}
        & \frac{\partial{T}^{\alpha\beta}}{\partial{\xi}^\alpha}+\Gamma^\alpha_{\alpha\lambda}T^{\lambda\beta}+\Gamma^\beta_{\alpha\lambda}T^{\alpha\lambda}-b^\beta_\alpha Q^\alpha+q^\beta=0, \ \ \  \beta=1,2 \\
        & T^{\alpha\beta}b_{\alpha\beta}-Q^\alpha_{|\alpha}+q^n=0.
        \end{aligned}
    \end{equation}
    The non-dimensional numbers in
    equations (\ref{a1}-\ref{a2}) are obtained using the radius of the cell $R$ as length scale, the shear rate $1/\dot{\gamma}$ as time scale, thus coupling membrane deformation and flow dynamics, $\rho_o\dot{\gamma}^2R^3$ as reference surface shear modulus and $\rho_o\dot{\gamma}^2R^5$ for the bending stiffness. 
    The  Weber number can be re-written as $We=Re \cdot Ca$ where $Ca=\mu_o R \dot{\gamma}/G_s$ is the Capillary number. In the present study, we shall consider  different stiffnesses of the cell membrane by varying the Capillary number and keeping the Reynolds number constant.
    
%    \noindent Finally, we note that for numerical purposes the space- and time-dependent viscosity field is given by an indicator function $I({\bf x}, t)$ 
%    \begin{equation}
%        \mu(\textbf{x},t)=\mu_o\left[1+(\mu^\ast-1)I(\textbf{x},t)\right].
%    \end{equation}
%%
%    \noindent Here we follow  Unverdi and Tryggvason \cite{unverdi1992front} for the definition of the indicator function as the solution to the following Poisson equation
%    \begin{equation}
%        \nabla^2 I = \nabla \cdot \textbf{G}    
%    \end{equation}
%    \noindent where the Green's function $\textbf{G}=\int_{\Sigma} \delta(X-x) \textbf{n} \ \mathrm{d}s$, and $\textbf{n}$ is the unit normal to the cell surface. 
%    \textcolor{red}{not clear. What is $\Sigma$, how is delta defined and so on...shall we move this part to the numerical section?}
%    \textcolor{green}{yes it's better to move it to numerical method like Kim and remove $\Sigma$}
%    Using the Dirac delta function in G makes the indicator function smoother near the boundary \cite{kim2015inertial}. Such indicator function is similar to the regularised Heaviside function used in the levelset framework.

\section{Numerical methods}
\label{sec: numerical methods}

    \subsection{The Navier-Stokes solver}\label{sec:NS}

    The Navier-Stokes equations are discretized using a staggered uniform grid to prevent checkerboard numerical instability, while the time integration relies on the classical projection method \cite{chorin1968numerical}. This method is a three-step procedure: first, a non-solenoidal velocity field ${\bf u}^*$ is computed as
        \begin{equation}
        \frac{{\bf u}^* - {\bf u}^{n}}{\Delta t} = {\bf RHS}({\bf u}, p),
    \end{equation}
    where ${\bf RHS}({\bf u}, p)$ is the right-hand side of the discretised Navier-Stokes equations and contains the fluid-structure interaction (FSI) forces. In a second step, the pressure field is obtained as the solution to the following Poisson equation
        \begin{equation}
        \nabla^2 p = -\frac{1}{\Delta t} \nabla \cdot {\bf u}^*.
        \label{eq: Poisson equation}
    \end{equation}
    Finally, the corrected velocity field ${\bf u}^{n+1}$ is obtained as
    \begin{equation}
        {\bf u}^{n+1} = {\bf u}^* + \Delta t \nabla p,
    \end{equation}
    where $\nabla p$ is the pressure gradient required for the velocity field ${\bf u}^{n+1}$ to be divergence-free. Second-order central differences are used for the spatial discretization of the convective terms, while their temporal integration relies on the Adams-Bashforth explicit method.
    
    As we allow for viscosity contrast between the fluid inside and outside the cells, the classical Fast Fourier spectral method cannot be readily used to evaluate the diffusive term ${\bf Du} = \nabla \cdot \left( \mu \left[ \nabla {\bf u} + \nabla {\bf u}^T \right] \right)$. Indeed, the viscosity field being a function of space, this operator cannot be reduced to a constant coefficient Laplace operator. However, Dodd \& Ferrante \cite{dodd2014fast} have recently introduced a splitting operator technique able to overcome this drawback. Though it has initially been derived for the pressure Poisson equation, this splitting approach can easily be extended to the Helmholtz equation resulting from an implicit (or semi-implicit) integration of the diffusive terms,
    ~
    \begin{equation}
        ({\bf I} - \Delta t {\bf D}){\bf u}^{n+1} = {\bf RHS}^n
    \end{equation}
    where ${\bf I}$ is the identify matrix, and ${\bf RHS}^n$ the discretized right-hand side including the non-linear advection terms. Given the viscosity field
    \begin{equation}
        \begin{aligned}
            \mu^\ast({\bf x}) & =1 + \mu'({\bf x})
        \end{aligned}
    \end{equation}
    where $1$ is the constant part and $\mu'({\bf x})$ the space-varying component, the diffusive term ${\bf Du}$ can be re-written as
    \begin{equation}
        {\bf Du} = \underbrace{\frac{1}{Re} \nabla^2 {\bf u}}_{{\bf D}_1{\bf u}} + \overbrace{\frac{1}{Re}\nabla \cdot \left( \mu'({\bf x}) \left[ \nabla {\bf u} + \nabla {\bf u}^T \right] \right)}^{{\bf D}_2{\bf u}}.
    \end{equation}
    
    \noindent The constant coefficients operator ${\bf D}_1$ can then be treated implicitly while the variable coefficients operator ${\bf D}_2$ is treated explicitly. The resulting Helmholtz equation then reads
    
    \begin{equation}
         ({\bf I} - \Delta t {\bf D}_1){\bf u}^{n+1} = {\bf RHS}^n + \Delta t {\bf D}_2{\bf u}^n .
         \label{eq: Helmholtz equation}
    \end{equation}
    
    \noindent Since ${\bf D}_1$ is now a constant coefficient Laplace operator, equation \eqref{eq: Helmholtz equation} can be solved using a classical Helmholtz solver based on Fast Fourier transforms. Note that a similar Fast Fourier-based solver is used to solve the Poisson equation \eqref{eq: Poisson equation} for the pressure.

    \subsection{Membrane representation: Spherical harmonics}\label{sec:SH}
    
    The membrane shape has been modeled as linear piece-wise functions on triangular meshes by Pozrikidis \cite{pozrikidis1995finite}, Ramanujan \& Pozrikidis \cite{ramanujan1998deformation} and Li \& Sarkar \cite{li2008front} among others. The finite element method has also been employed by Walter {\it et al.} \cite{walter2010coupling} for its generality and versatility. Another interesting method is the global spectral method. Fourier spectral interpolation and spherical harmonics have been used for two-dimensional \cite{freund2007leukocyte} and three-dimensional simulations \cite{kessler2008swinging,zhao2010spectral}. Here, we follow the approach of Zhao {\it et al.} \cite{zhao2010spectral} , previously implemented in \cite{zhu2015motion,zhu2014microfluidic,zhu2015dynamics,rorai2015motion}. This is briefly outlined below.
    
    The capsule surface is mapped onto the surface of the unit reference sphere $S^2$,  using the angles in spherical coordinates $(\theta,\phi)$ for the parametrization. The parameter space $\{(\theta,\phi)\mid 0\leq\theta\leq\pi, 0\leq\phi\leq2\pi\} $ is discretized by a quadrilateral grid using Gauss-Legendre (GL) quadrature intervals in $\theta$ and uniform spacing in the $\phi$ direction. All other surface quantities are stored on the same mesh, {\it i.e.}\ the grid is co-located. The surface coordinates $x(\theta,\phi)$ are expressed by a truncated series of spherical harmonic functions,
    \begin{equation}
        x(\theta,\phi)=\sum_{n=0}^{N_{SH}-1}\sum_{m=0}^{n} \bar{P}^m_n(\cos{\theta})(a_{nm}\cos{m\phi}+b_{nm}\sin{m\phi}),    
    \end{equation}
yielding $N^2_{SH}$ spherical harmonic modes. The corresponding normalized Legendre polynomials are given by
    \begin{equation}
    \bar{P}^m_n(x)=\frac{1}{2^nn!}\sqrt{\frac{(2n+1)(n-m)!}{2(n+m)!}}(1-x^2)^{{m}/{2}}\frac{d^{n+m}}{dx^{n+m}}(x^2-1)^n.    
    \end{equation}

    The SPHEREPACK library \cite{adams1999spherepack,swarztrauber2000generalized} is employed for the forward and backward transformations. To deal with aliasing errors arising due to the nonlinearities in the membrane equations (products, roots and inverse operations needed to calculate the geometric quantities), we use an approximate de-aliasing by performing the nonlinear operations on $M_{SH}>N_{SH}$ points and filtering the result back to $N_{SH}$ points. A detailed discussion on this issue is provided in Freund \& Zhao \cite{freund2010high}. For most of the results presented in the present study, 576 spherical harmonics with 24 modes are used to define the cell shape.
    
    Considereing different viscosity inside and outside the cell, a space and time dependent viscosity field is defined by an indicator function $I({\bf x}, t)$ related to the membrane location,
    \begin{equation}
            \mu^\ast({\bf x})  = (1 - I({\bf x})) + I({\bf x})\lambda                  . 
    \end{equation}
    \noindent Here we follow  Unverdi and Tryggvason \cite{unverdi1992front} for the definition of the indicator function as the solution to the following Poisson equation
    \begin{equation}
        \nabla^2 I = \nabla \cdot \textbf{G}    
    \end{equation}
    \noindent where the Green's function $\textbf{G}=\int \delta(X-x) \textbf{n} \ \mathrm{d}s$, and $\textbf{n}$ is the unit normal vector to the cell surface. 
    Using the smooth Dirac delta function introduced below in the computation of G makes the indicator function smoother near the boundary \cite{kim2015inertial}. Such indicator function is similar to the regularised Heaviside function used in the levelset framework.

    \subsection{Immersed boundary method}
        \subsubsection{Immersed boundary method for deformable partices}
    
    The immersed boundary method \cite{peskin2002immersed} is commonly used to solve fluid-solid interaction problems. In this method, two distinct sets of grid points are used: ({\it i}) an Eulerian mesh to solve fluid flow, see section \ref{sec:NS}, and ({\it ii}) a Lagrangian mesh for solving the particle motion, section \ref{sec:SH}. In the present  low Reynolds number framework, we start from the original approach of Peskin \cite{peskin2002immersed}. At each time step, the fluid velocity defined on the Eulerian mesh is first interpolated onto the Lagrangian mesh,
    \begin{equation}
    \textbf{U}_{ib}(X,t)=\int_\Omega\textbf{u}(x,t)\delta(\textbf{X}-\textbf{x})dx,
    \end{equation}
    where $x$ and $X$ are the Eulerian and Lagrangian coordinates and $\delta$ is a smooth Dirac delta function, here that proposed in Ref.\ \cite{roma1999adaptive}. The elastic force per area $\textbf{q}$ and surface normal vectors ${\bf n}$ are then computed from the membrane equations described above. As next step, the normal vectors are used to compute the indicator function $I({\bf x}, t)$ on the Eulerian mesh. The force is then spread to Eulerian mesh and added to the momentum equations as
    
    \begin{equation} \label{sp}
    \textbf{f}(\textbf{x},t)=\int\textbf{q}(\textbf{X},t)\delta(\textbf{x-\textbf{X}})ds.
    \end{equation}

    \noindent Thereafter, the positions of the Lagrangian points are updated according to
    
    \begin{equation} \label{eq5}
    \textbf{X}^{n+1}=\textbf{X}^n+\int^t_0 \textbf{U}_{ib} dt.
    \end{equation}
    
\begin{figure}
    \centering
    \includegraphics[width=.9\columnwidth]{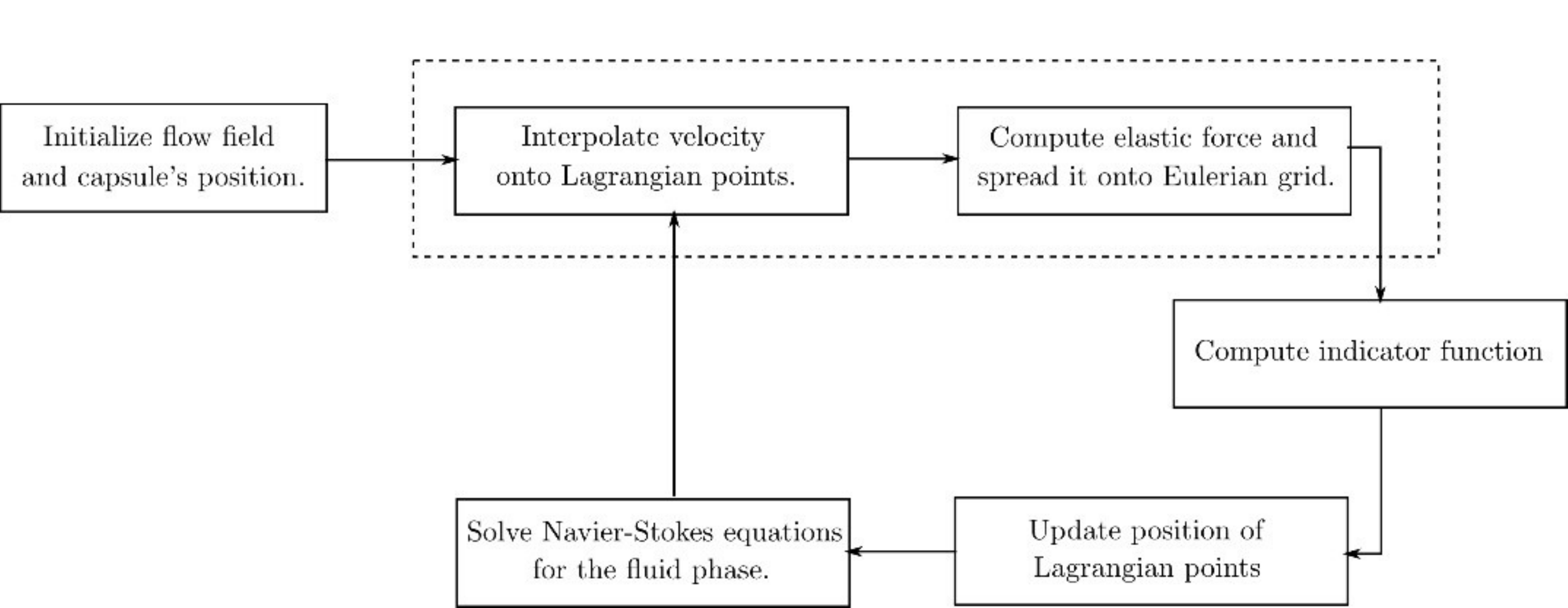}
    \put(-230,153){\footnotesize Immersed boundary method}
    \caption{A flowchart of the computational procedure used in present study}
    \label{flowchart}
    \end{figure}

    Note that equation (\ref{eq5}) assumes an over-damped regime, i.e.\ the Lagrangian points go to their equilibrium position immediately after the FSI force is applied. Finally, the fluid flow is solved in the Eulerian framework as explained in section \ref{sec:NS}. A flowchart for computational procedure at each iteration is depicted in figure \ref{flowchart} .

    \medskip
    
    The method described above is not particularly efficient when the Reynolds number increases since it requires very small time steps, which  increases the computational time. At moderate and high Reynolds numbers, a modification of the method by Kim {\it et al.} \cite{kim2015inertial}, is employed to be consistent with the assumption of inertialess membrane. In our approach, in addition to the  Lagrangian coordinates $\bf{X}$, we introduce the additional immersed boundary points $\bf{X}_{ib}$ whose motion is governed by equation (\ref{eq5}).  Since the total force exerted on each element on the membrane surface is equal to the difference between its acceleration and the acceleration of fluid element at the same location,
    the motion of the real Lagrangian points is governed by
    
    \begin{equation} \label{eq6}
    \rho_{os}\frac{\partial^2{\textbf{X}}}{\partial{t}^2}=\rho_{os}\frac{\partial^2{\textbf{X}_{ib}}}{\partial{t}^2}+\textbf{F}_e-\textbf{F}_{FSI}+\textbf{F}_A
    \end{equation}
    
    \noindent where $\rho_{os}$ is the surface density of the base fluid. The two sets of Lagrangian points, $\textbf{X}$ and $\textbf{X}_{ib}$, are connected to each other by a spring and damper, {\it i.e.}\ a fluid-solid interaction force $\textbf{F}_{FSI}$ computed using the following feedback law
    
    \begin{equation} \label{eq7}
        \textbf{F}_{FSI}=-\kappa \left[(X_{ib}-X)+\Delta t(\textbf{U}_{ib}-\textbf{U})\right].
    \end{equation}
    
    The final modified procedure is therefore as follows. At each time step, we first compute $\textbf{U}_{ib}$ and the fluid-solid interaction force $\textbf{F}_{FSI}$  from equation (\ref{eq7}). The indicator function $I({\bf x}, t)$ is then computed to identify the interior of the cell and impose viscosity contrasts, and the momentum equation solved to obtain the flow velocity ${\bf u}$. Finally, the positions of the Lagrangian points $\bf{X}$ are updated using equation (\ref{eq6}). This additional equation is made non-dimensional as above, in particular with $\rho_{o}R$ for the surface density of the base fluid and $\rho_{o}R^2\dot{\gamma}^2$ for the elastic and fluid-solid interaction forces per unit area. For completeness, we report the non-dimensional form of equation (\ref{eq6}),
    
    \begin{equation}\label{eq8}
    d^\ast\frac{\partial^2{\textbf{X}}}{\partial{t}^2}=d^\ast\frac{\partial^2{\textbf{X}_{ib}}}{\partial{t}^2}+\textbf{F}_e-\textbf{F}_{FSI}+\textbf{F}_A
    \end{equation}

    \noindent where $d^\ast=d/R$ is ratio between the membrane thickness and initial radius of the cell, assumed in the present study to be $d^\ast=0.01$. In the above, $\textbf{F}_A$ is the penalty force used to enforce volume conservation, calculated as in \cite{kim2015inertial}:
    
    \begin{equation}
    \begin{aligned}
    &\textbf{F}_A=\Delta p \cdot \eta (\theta,\phi) \cdot \textbf{e}_n,\\ 
    &\Delta p=\frac{1}{\beta}\left(1-\frac{V}{V_0}\right)+\frac{1}{\beta}\int_0^t\left(1-\frac{V}{V_0}\right)dt^{\prime}.
    \end{aligned}
    \end{equation}
 Here $\Delta p$ represents the pressure generated by the volume change, $\eta (\theta,\phi)$ is the surface area of each element and $\textbf{e}_n$ the local unit normal vector. This force is also added to the elastic force $\textbf{q}$ before spreading it to the Eulerian mesh according to equation (\ref{sp}). 
    
    \subsubsection{Immersed boundary method for rigid particles}

    In the present study, two different immersed boundary methods are considered to simulate the stiff nucleus. First, the  method described above is used with high surface shear modulus for the inner capsule. In the second approach we follow the implementation by  Breugem \cite{breugem2012second}, which has been widely used in the framework of rigid particles, see e.g.\ \cite{lashgari2014laminar}. In this method, a moving Lagrangian mesh is adopted to impose no-slip and no-penetration on the surface of a rigid object. The numerical procedure is as follows: first, the prediction velocity $\textbf{u}^\ast$ is computed from the Navier-Stokes equations neglecting the fluid-solid interaction force. This fluid velocity is then interpolated onto the Lagrangian mesh $(\textbf{U}^\ast)$ and the fluid-solid interaction force computed, based on the difference between the fluid and the solid body velocity at each Lagrangian point,
        \begin{equation}
        \textbf{F}_{FSI}=\frac{\textbf{U}_P-\textbf{U}^\ast}{\Delta t}.
    \end{equation}
     This force is spread to the Eulerian grid and the second prediction velocity $\textbf{u}^{\ast\ast}$ obtained by solving the Navier-Stokes equations with the fluid-solid interaction force. The divergence-free constraint is then imposed on the velocity field by solving the pressure Poisson equation and correcting the velocity field appropriately. Finally, the total force and torque on each particle is computed, and the translational and rotational velocities of the particle obtained by integrating the Newton-Euler equations. Readers are also referred to \cite{uhlmann2005immersed} for further details.     
  
    \subsection{Notes on the parallelisation and implementation}
    
    The Eulerian mesh is decomposed using a 2D-pencil domain decomposition in the streamwise and spanwise direction. For that purpose, the library 2DECOMP \& FFT \cite{li20102decomp} is used. The same library handles all of the transpose operations required for the Helmholtz and Poisson solvers based on the fast Fourier transforms. Regarding the parallelization of each particle/capsule, each processor can either be {\it master} or {\it slave}. A processor is labelled {\it master} if it contains most of the Lagrangian points describing the given particle, while those containing only some of these points are labelled as {\it slaves}. The rest of the processors do not have any rule for the considered particle. Only the {\it master} processor has the information of the particle ({\it e.g.} Lagrangian points and their velocities) in its memory, though the slaves can access it for interpolation and spreading operations, which might require information from the neighbours. For the particle equations, the master is responsible for all the numerical procedures, e.g.\ transformation using spherical harmonics. Such parallelization saves memory usage but requires communication between cores at each time step. \\   
%    \textcolor{red}{Regarding boundary conditions, We have no-slip boundary condition at walls and the flow is considered to be periodic in steamwise and spanwise directions }\\
 % \medskip
    In order to obtain accurate results, the density of Eulerian and Lagrangian grid points should be similar. In some cases, it is nonetheless necessary to have a very fine Eulerian mesh, thus requiring very fine Lagrangian mesh. However, since the spherical harmonic calculations are costly, the overall computational time increases significantly. To make the code more efficient, two different sets of Lagrangian points are therefore considered: forcing points that are used for interpolation-spreading and the spherical-harmonic points that are used to define the shape of the cell and the elastic stresses. While the density of the forcing points has to be similar to that of the Eulerian points, fewer points are required for the spherical harmonics representation of the cell, especially in the case of stiffer membranes deforming less. At each time step, before computing the elastic forces, the spherical-harmonic points are obtained using spectral interpolation. These points are then used to compute the elastic forces and surface normal vectors. Once computed, the elastic forces and surface normal vectors are interpolated onto the finer mesh so that the ealstic forces are spread on the Eulerian mesh. All spectral interpolations are done with the SPHEREPACK library \cite{adams1999spherepack,swarztrauber2000generalized}.

    \section{Validations}
    \label{sec: validations}
    
    In order to validate the current implementation, two different cases are considered for a simple capsule: deformation of a capsule in shear flow at low Reynolds number and equilibrium position of a capsule in Poiseuille flow at finite Reynolds number. The domain is a box with the size $10\times10\times10$ in units of cell radius, and the cell is located at its center. The Eulerian grid is $128^3$ whereas for the Lagrangian mesh $24\times48$ points are chosen in the latitudinal and longitudinal directions respectively.
The non-dimensional bending stiffness $\frac{G_B}{\rho_o R^5\dot{\gamma}^2}$ is zero unless otherwise mentioned. The dealiasing ratio is kept at $M_{SH}/N_{SH}=2$.    

   A grid independence study has been carried out for $Ca=0.6$ by increasing the number of the Eulerian grid points and of spherical-harmonic grids by a factor of 1.5. In this case, the change in the deformation parameter is found to be less than $2\%$. 
   In addition, we also increased the box size by a factor 1.5 and measured a difference in the deformation parameter below $0.3\%$. 
   To check time step independency of the results, we decreased the time step by $50\%$ and obtain a change in the deformation parameter of only $0.002\%$. 
   Finally, 
   the effect of the ratio between the forcing points and the number of spherical harmonics describing the membrane deformation is investigated using 1152 force points. When the ratio between the two is  $4$, the difference measured by the deformation parameter is less than $0.5\%$ with respect to the case with  ratio $1$.
Increasing  the ratio from $1$ to $2.25$, the change is of about $0.1\%$ . 
Here we used the more demanding ratio of 1 to be sure to capture all features of the cell shape for large deformations. 
   
    \subsection{Single cell in shear flow}

\begin{figure}
    \centering
    \includegraphics[width=.8\columnwidth]{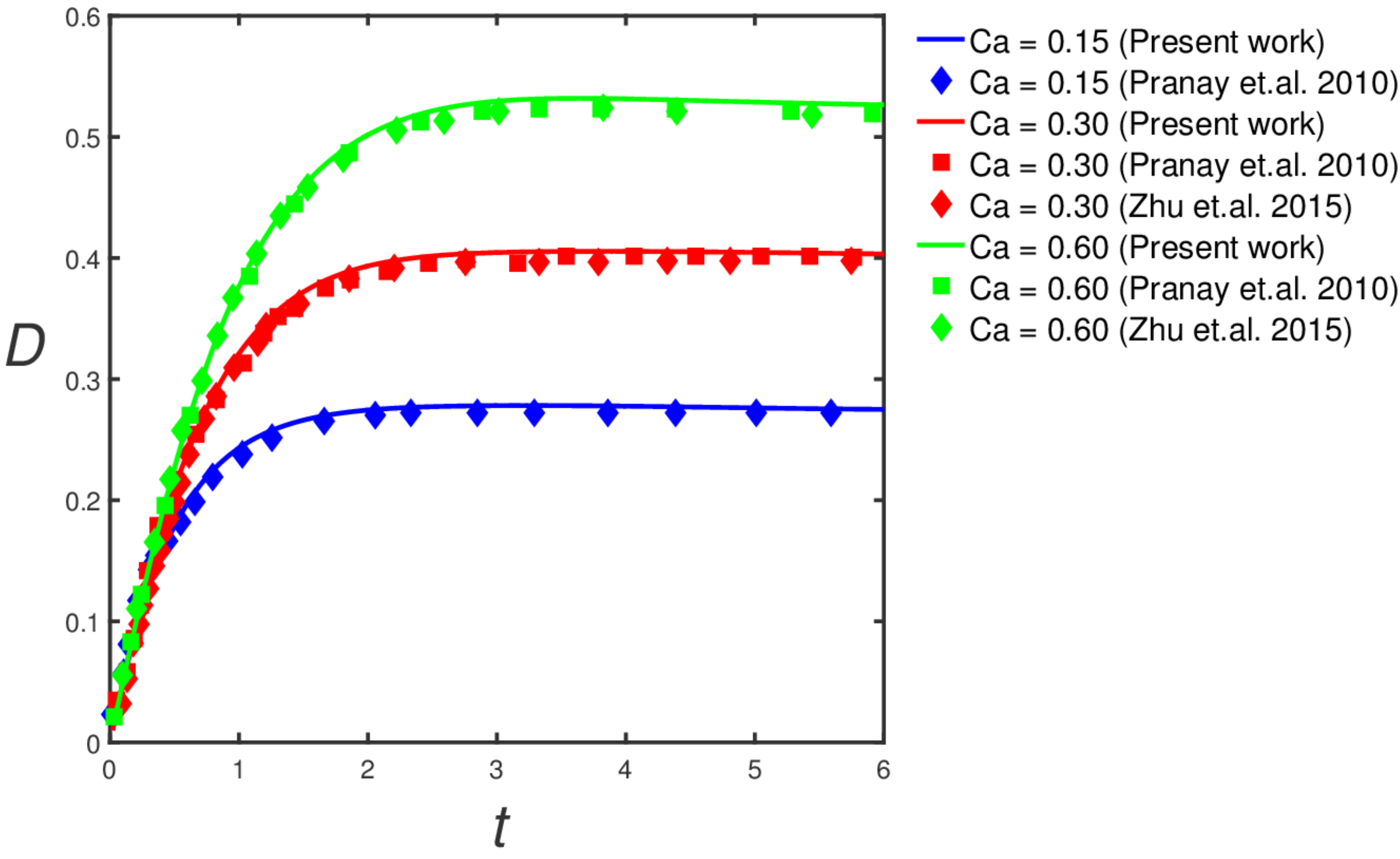}
    \caption{Deformation parameter versus time for an initially spherical cell in shear flow for different capillary numbers. The results are compared with those in Ref. \cite{pranay2010pair} and \cite{zhu2015motion}.}
    \label{f1}
    \end{figure}

    In this section, the deformation of a cell in a simple shear flow obtained with the present implementation is compared to the results by Pranay {\it et al.} \cite{pranay2010pair} and Zhu {\it et al.} \cite{zhu2015motion}. These authors have used boundary integral approaches to solve for the Stokes flow, so the Reynolds number is chosen here small enough, $Re=0.1$, to ensure low inertia of the flow. The cell deformation is quantified by the parameter
    \begin{equation}
    D=\frac{\left|l_2-l_1\right|}{l_2+l_1}    
    \end{equation}
    \noindent where $l_1$ and $l_2$ denote the major and minor semi-axes of the equivalent ellipsoid in the middle plane, respectively. The inertia tensor of the cell is used to obtain the equivalent ellipsoid as in \cite{ramanujan1998deformation}. As shown in figure \ref{f1}, a good agreement is obtained with the different implementations.

    \subsection{Capsule in channel flow at finite inertia}

\begin{figure}
    \centering
    \subfigure[]{\includegraphics[width=.45\columnwidth]{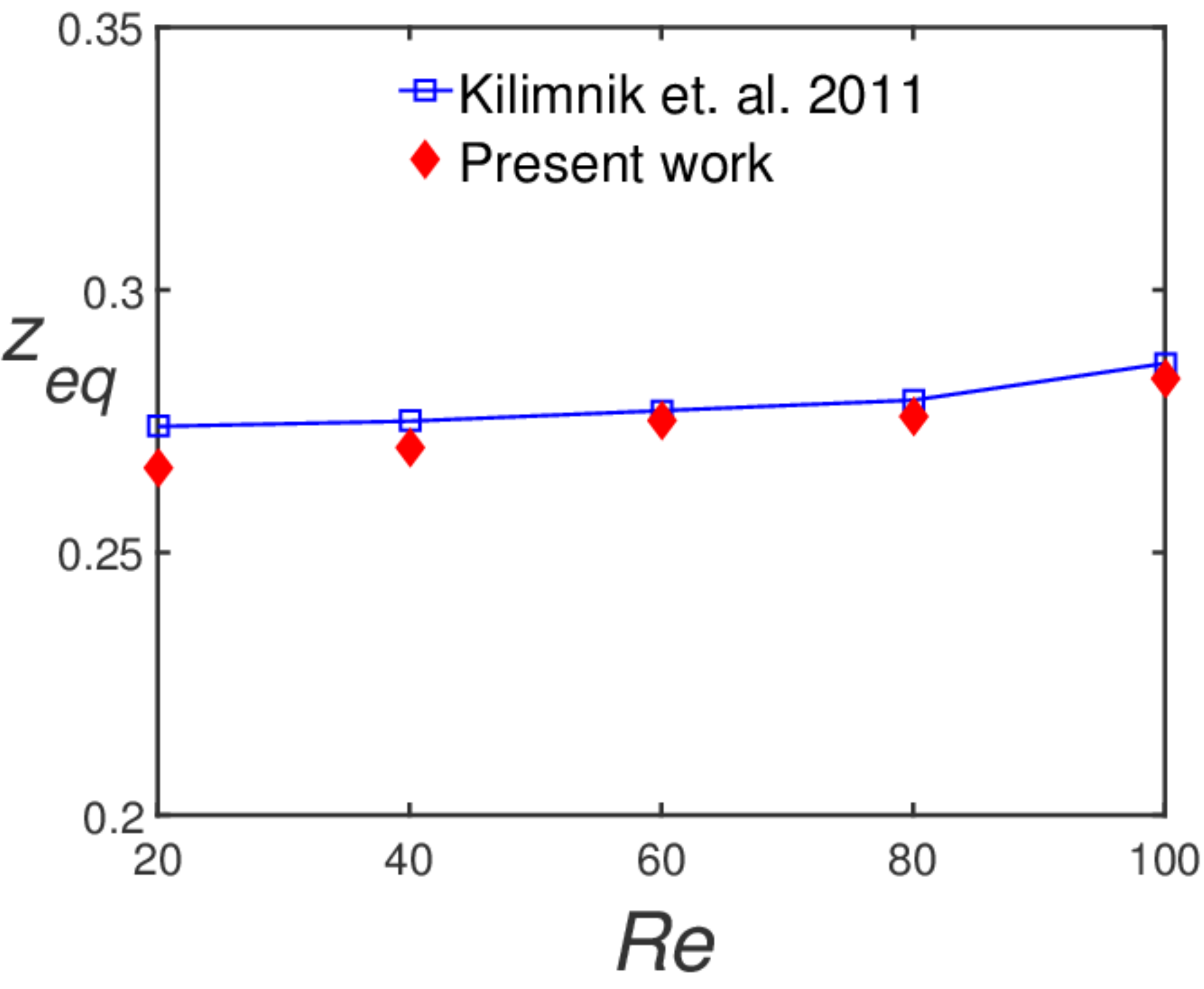}}
    \subfigure[]{\includegraphics[width=.45\columnwidth]{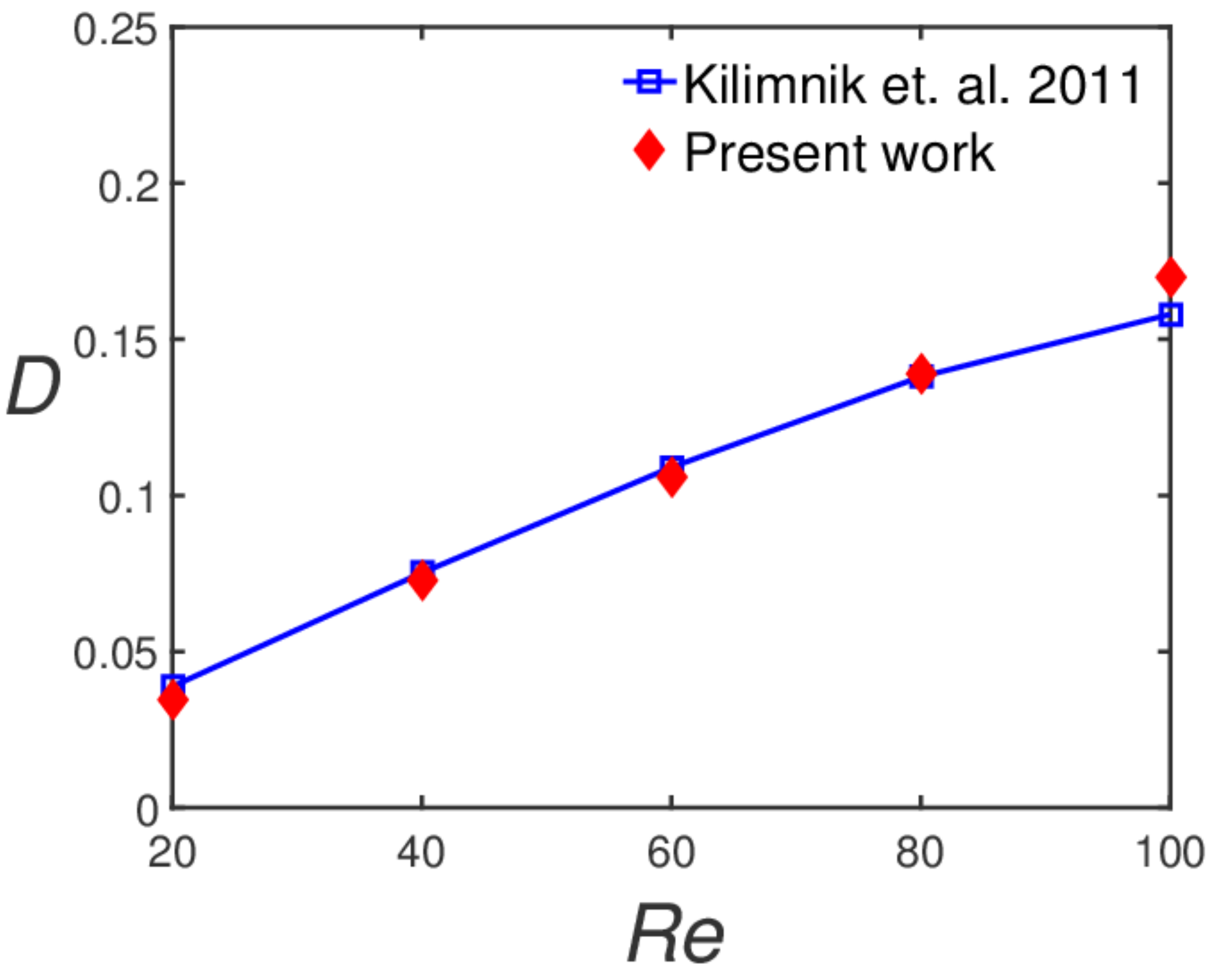}}
    \caption{(a) Eqilibrium position of an initially spherical cell transported in a pressure driven channel flow as a function of Reynolds number. (b) Deformation parameter of an initially spherical cell as a function of the Reynolds number. In both figures, the Capillary and bending stifness are set to $Ca=0.174$ and $B=0.02$, respectively.}
    \label{f2}
\end{figure}

    The second validation case is provided by the equilibrium position of a cell in Poiseuille flow at finite Reynolds numbers. Indeed, after some transients, a single capsule reaches a steady state characterised by a constant wall-normal position, constant velocity and deformation. This equilibrium position is located between the wall and the channel centerline and depends on both the Reynolds and Capillary numbers \cite{kilimnik2011inertial}. In the present validation, the Capillary number is set to $Ca = 0.174$ and the Reynolds number is varied. Note that since the present bending model is different from that in the work of Kilimnik et al \cite{kilimnik2011inertial}, we chose the value of the bending stiffness that best fit their data, B=.02. The dependence of the wall-normal position and deformation of the capsule on the Reynolds number are reported on figures \ref{f2}(a) and \ref{f2}(b), respectively. As shown by these figures, a good agreement with the results in \cite{kilimnik2011inertial} is obtained. In figure \ref{f2}(b), the difference appearing at $Re=100$ may be attributed to the model considered in \cite{kilimnik2011inertial}. These authors used a Hookean law with a lattice-spring model (LSM) where the membrane has a finite thickness whereas in present work we use the hyperelastic neo-Hookian model with a infinitely thin membrane. At small Reynolds numbers, when the deformation is relatively small, the last term on the right hand side of equation (\ref{a3}) vanishes and our model practically reduces to a Hookean law: the two methods give thus very similar results. In the case of large deformations, the term introducing non-linearity in the constitutive model is not negligible any longer and the results start to deviate from each other.   

\section{Results}
\label{sec: results}

    \subsection{Stiff nucleus}
    
    As a starting point, the two-membrane model, based on the original IBM by Peskin \cite{peskin2002immersed}, is employed. In order to have a stiff nucleus, the capillary number of the nucleus is chosen to be 300 times smaller than that of the outer membrane. The cell is subject to homogeneous shear as in the validation cases presented above, {\it i.e.} same configuration and numerical parameters. 
    
\begin{figure}
\centering
\subfigure[]{\includegraphics[width=.45\columnwidth]{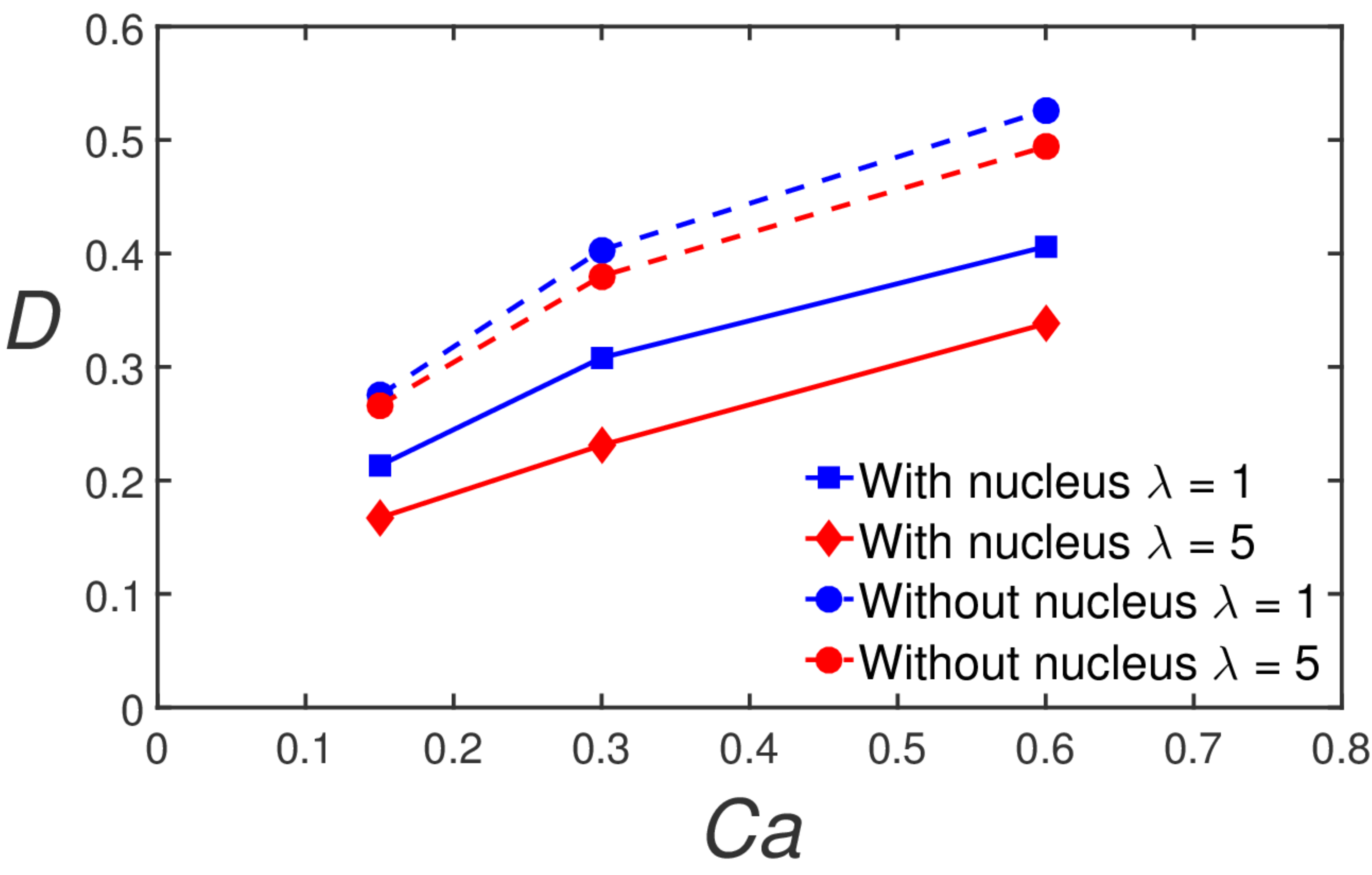}}
\subfigure[]{\includegraphics[width=.45\columnwidth]{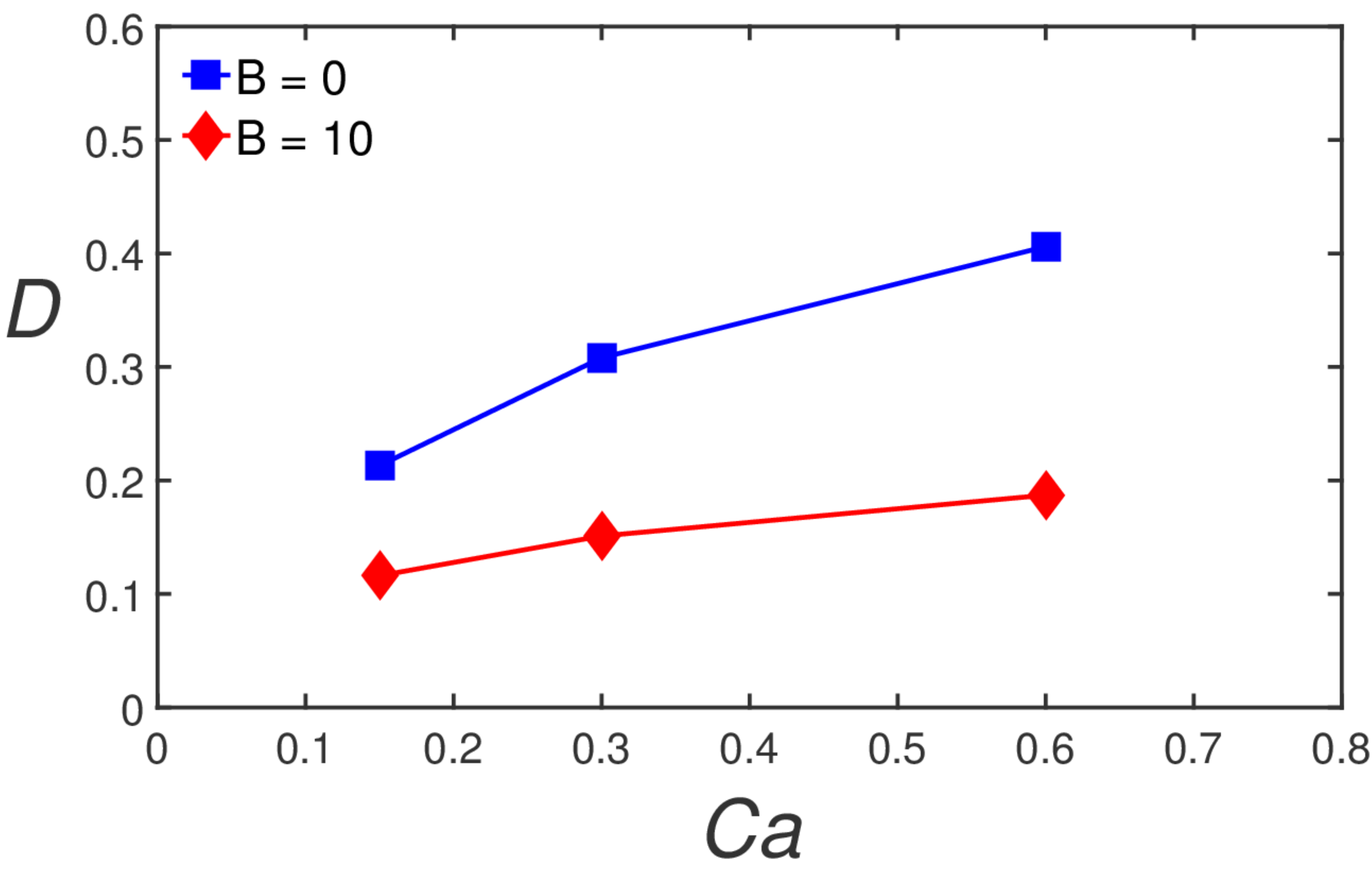}}
\caption{(a) Deformation parameter of initially spherical cells with and without a stiff nucleus in homogeneous shear versus the Capillary number for viscosity ratios $\lambda=1$ and 5 and no bending stiffness. (b) Cell deformation parameter versus the Capillary number for capsules with and without bending stiffness at viscosity ratio $\lambda=1$ and $Re=0.1$.}
\label{f4}
\end{figure}

   Figure \ref{f4} depicts the deformation parameter as a function of the Capillary number for viscosity ratios $\lambda=1$ and 5 in the absence of bending resistance, both for capsules with and without a stiff nucleus. Note that the deformation parameter is computed on the outer membrane, the inner one not being noticeably deformed. As shown in this figure, the presence of a nucleus reduces the deformation, and this reduction is larger the higher the Capillary number. The stiff nucleus reduces the outer membrane deformation since the minimum radius cannot be smaller than the radius of the nucleus. At higher Capillary numbers, the membrane would tend to deform more thus making the effect of the nucleus becomes more evident. It can be inferred from figure \ref{f4} that the deformation is smaller for the cases with a more viscous fluid between the outer membrane and the nucleus, $\lambda=5$. In this case, viscous forces appear to work together with elastic forces to reduce the cell deformation.
    
\begin{figure}
    \centering
    \includegraphics[width=.5\columnwidth]{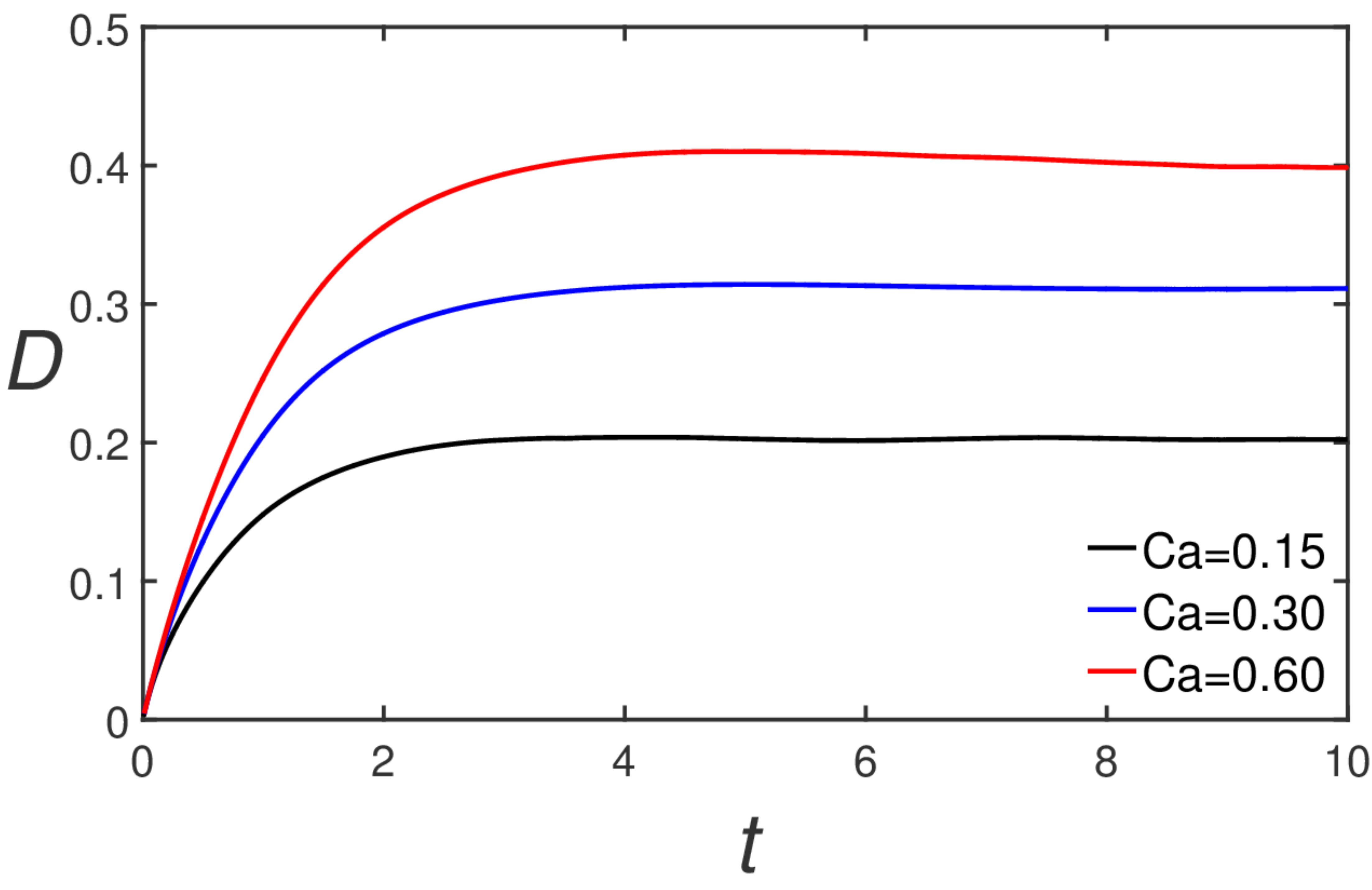}
    \caption{Time evolution of deformation parameter for an initially spherical cell with nucleus for different capillary numbers. Reynolds number $Re=0.1$, bending stiffness $B=0$.}
    \label{ca}
    \end{figure}

    %\textcolor{blue}{We need to check the data in  figure  \ref{ca}.}     
    The transient evolution of the deformation parameter to reach the final state is demonstrated in figure \ref{ca} for three different capillary numbers. The figure shows that larger capillary numbers require longer time to reach the final steady state. As for the steady state, the deformation is larger for higher capillary numbers.
  
\begin{figure}
    \centering 
    \subfigure{
    \includegraphics[scale=.15]{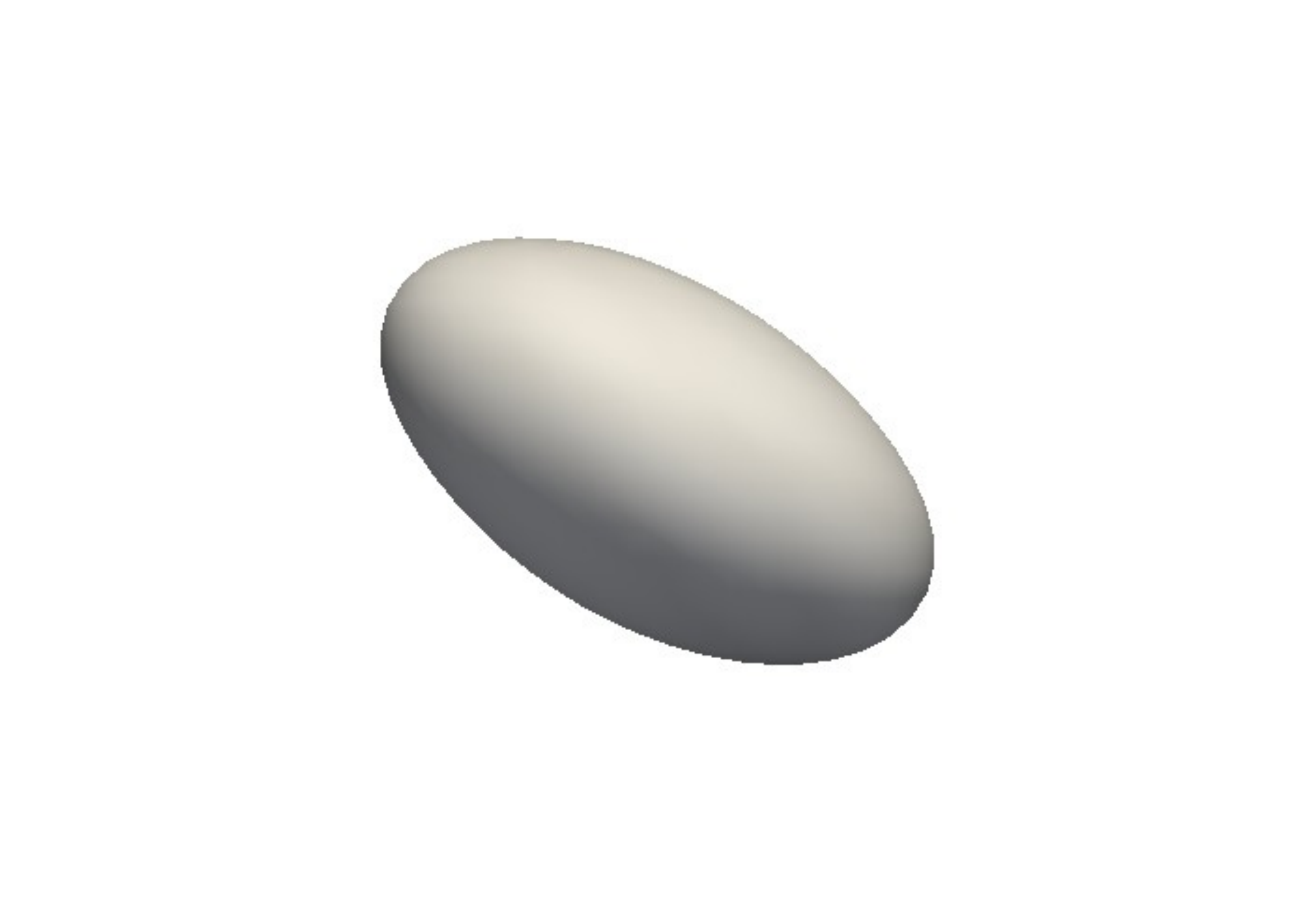}
    \label{sb11}
    \put(-120,33){\footnotesize $(i)$}
}
    \subfigure{
    \includegraphics[scale=.15]{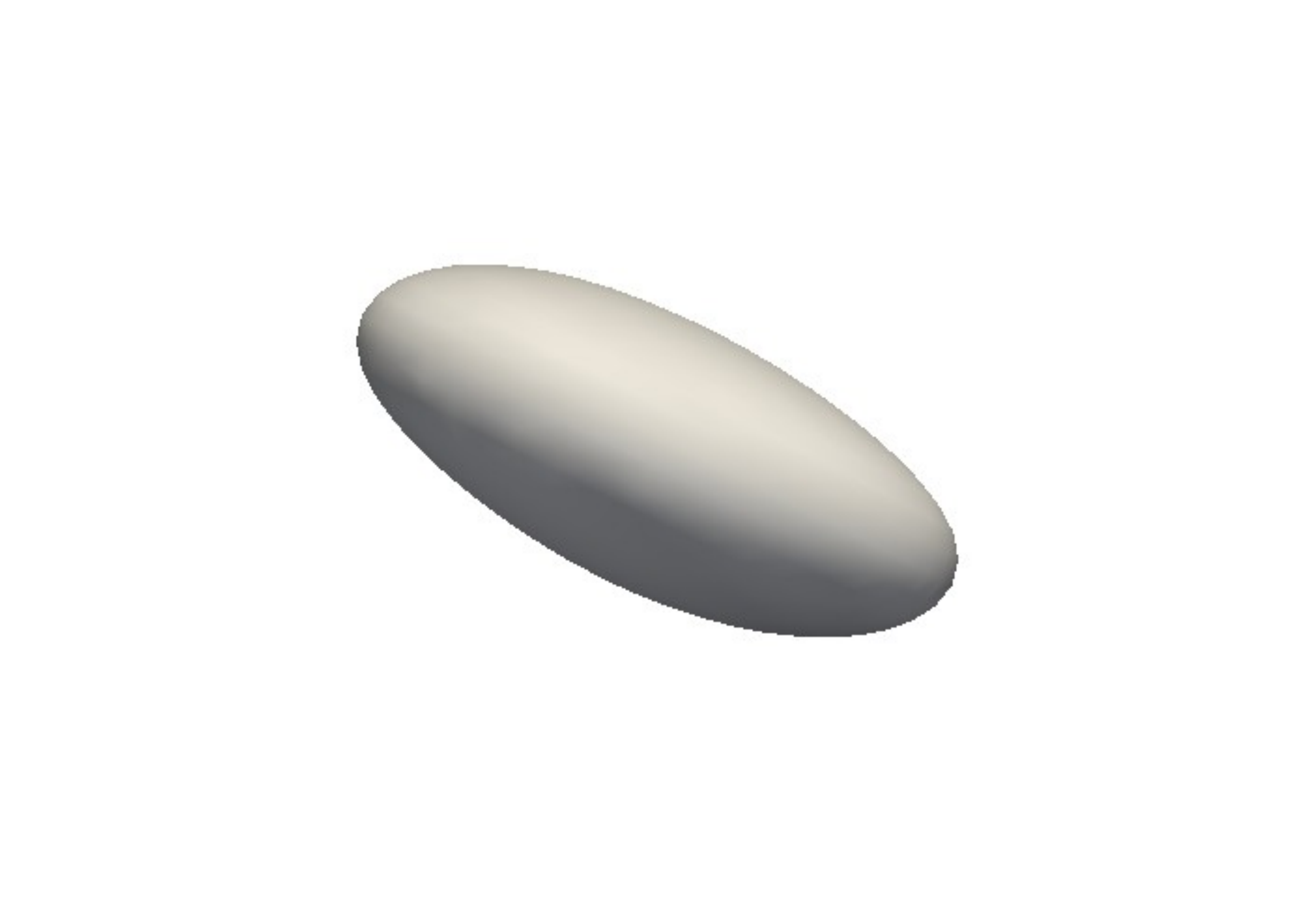}
    \label{sb12}
}
    \subfigure{
    \includegraphics[scale=.15]{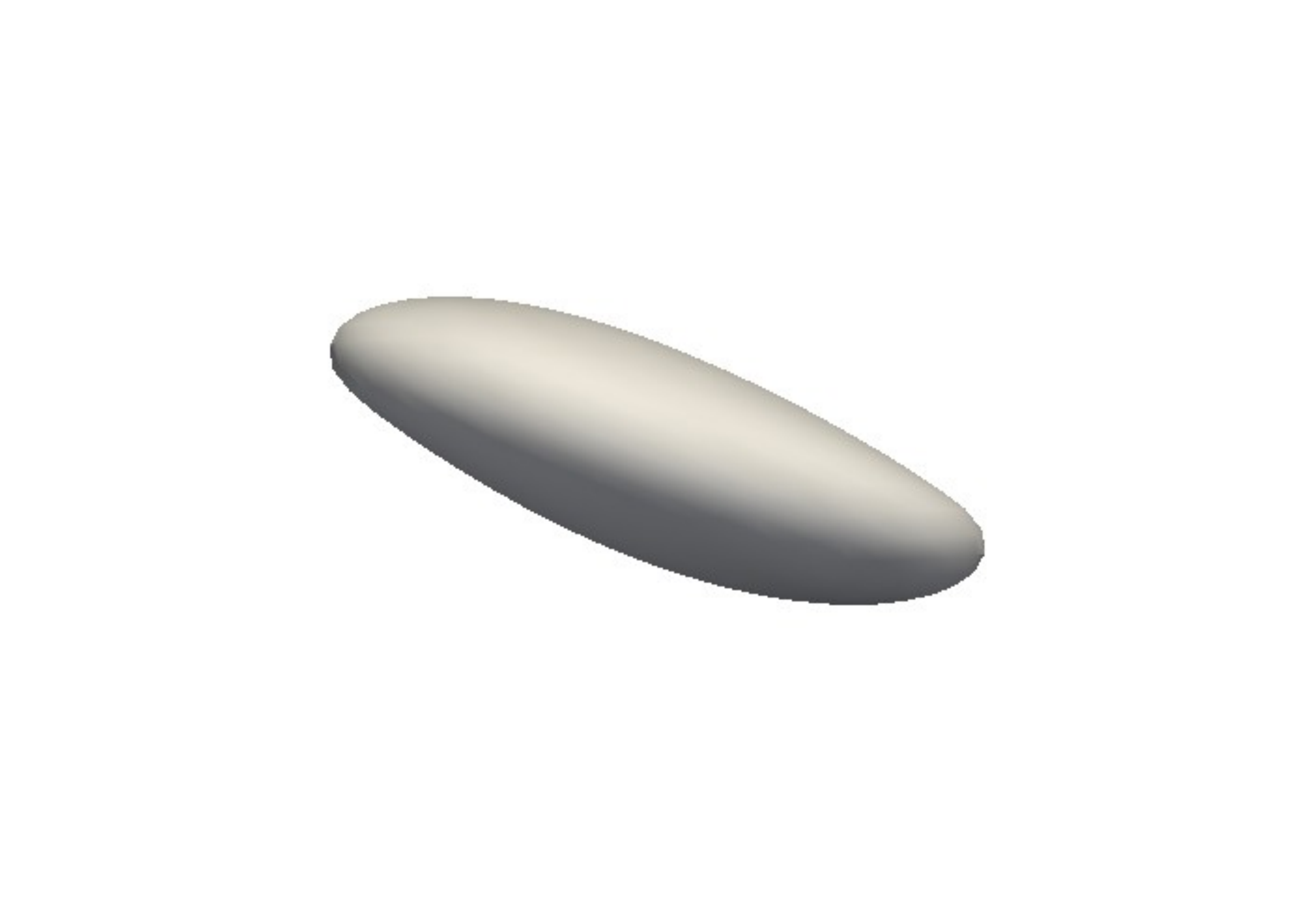}
    \label{sb13}
}\\
    \subfigure{
    \includegraphics[scale=.15]{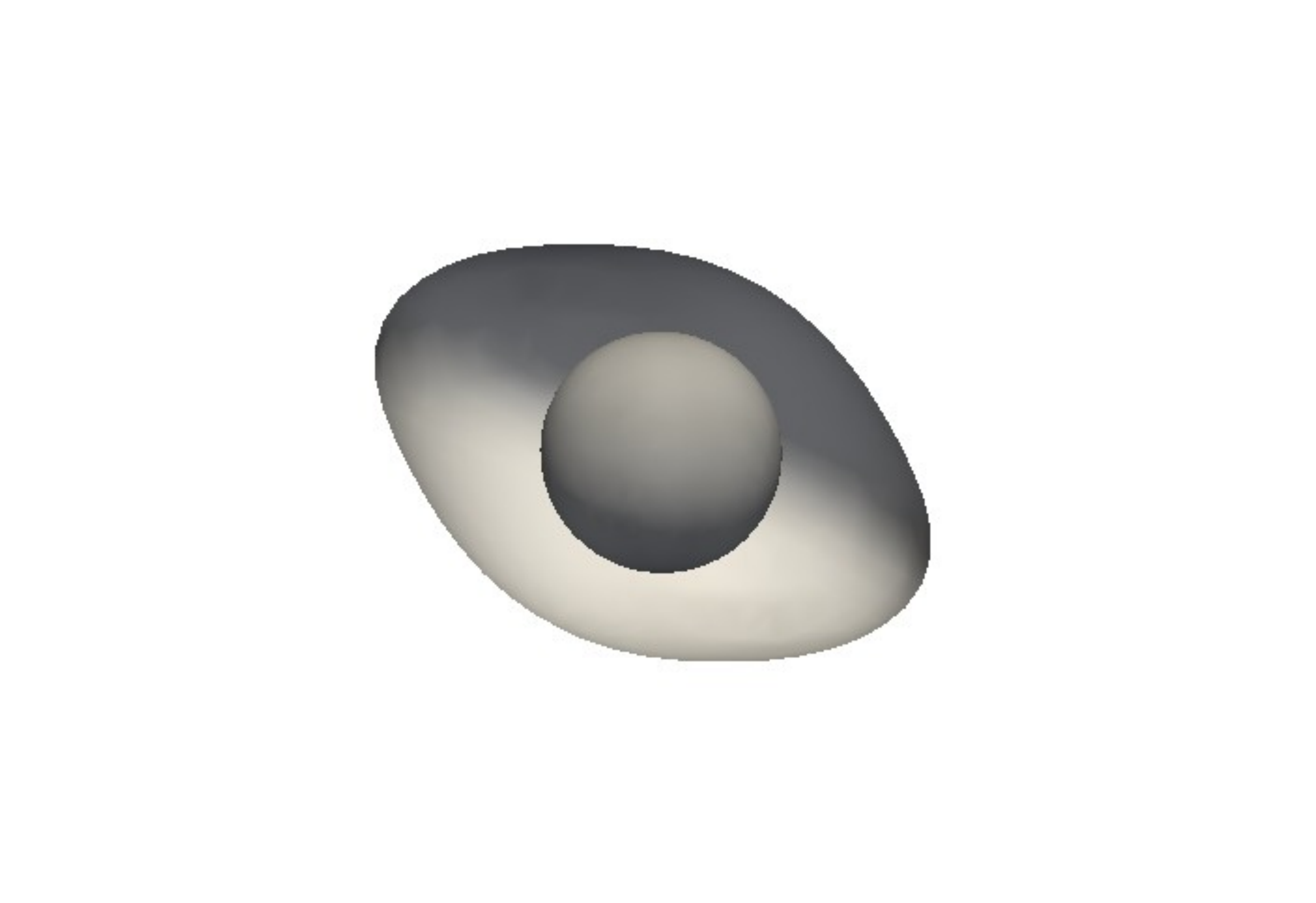}
    \label{sb21}
        \put(-120,33){\footnotesize $(ii)$}
}
    \subfigure{
    \includegraphics[scale=.15]{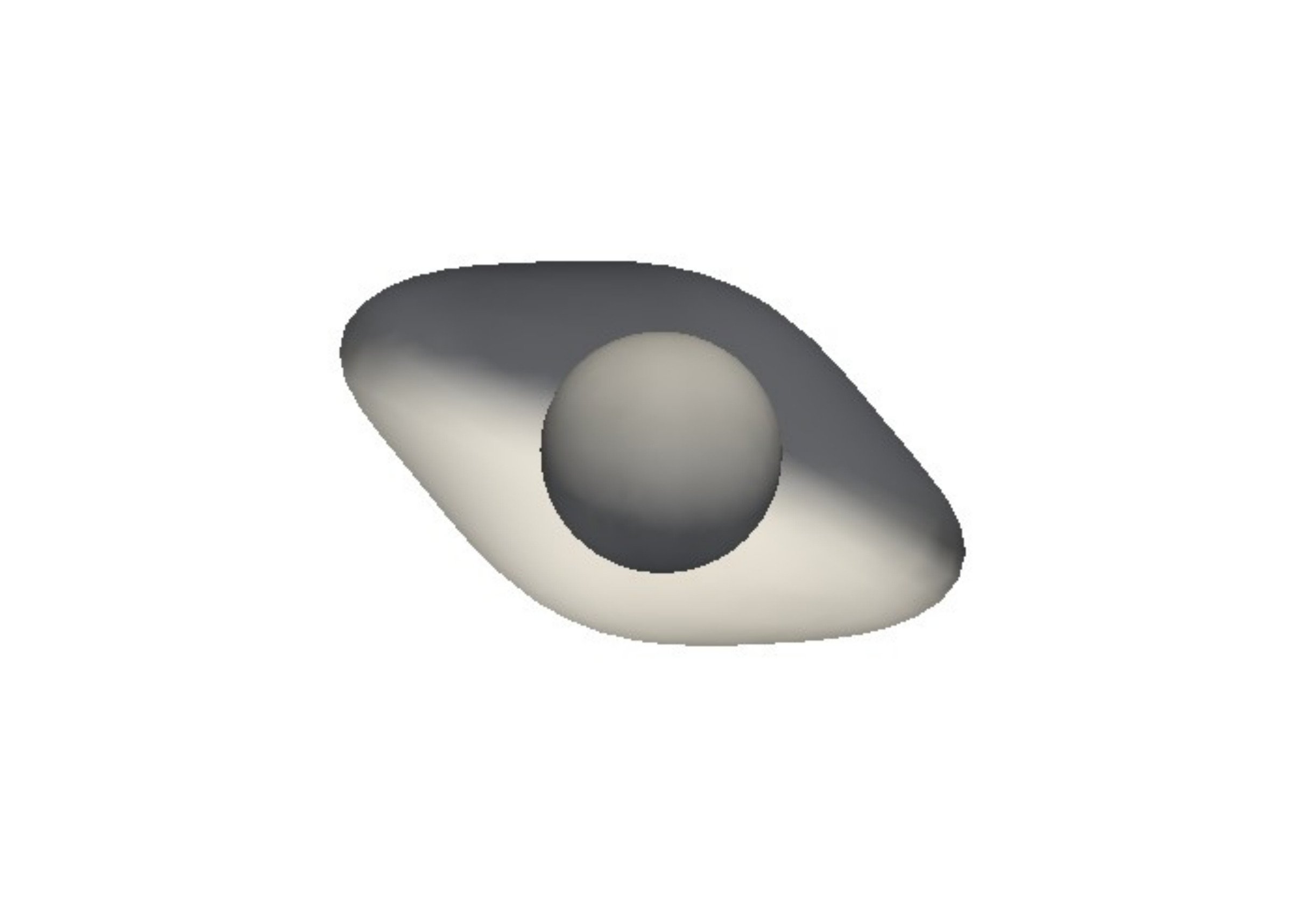}
    \label{sb22}
}
    \subfigure{
    \includegraphics[scale=.15]{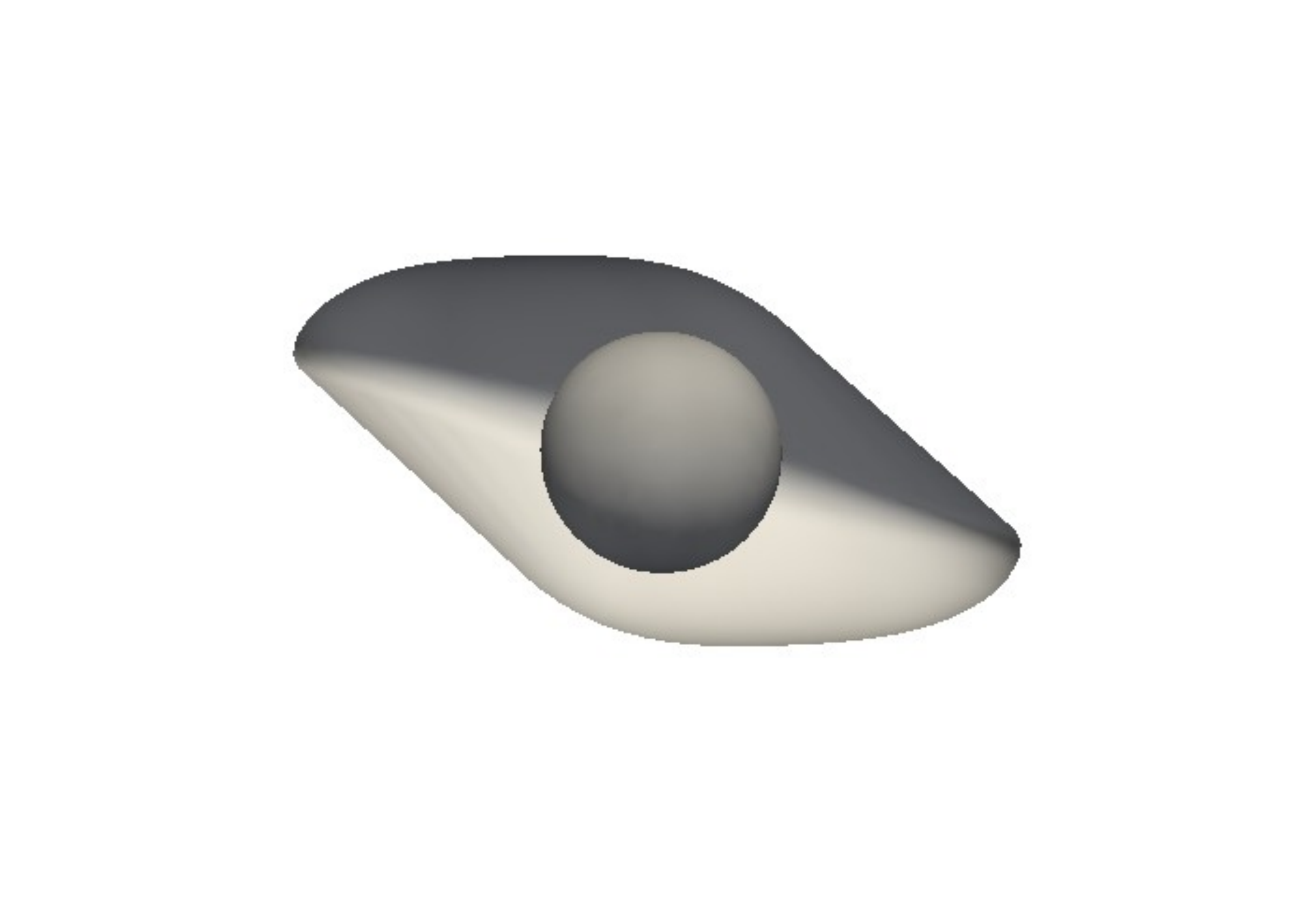}
    \label{sb23}
}\\
    \subfigure{
    \includegraphics[scale=.148]{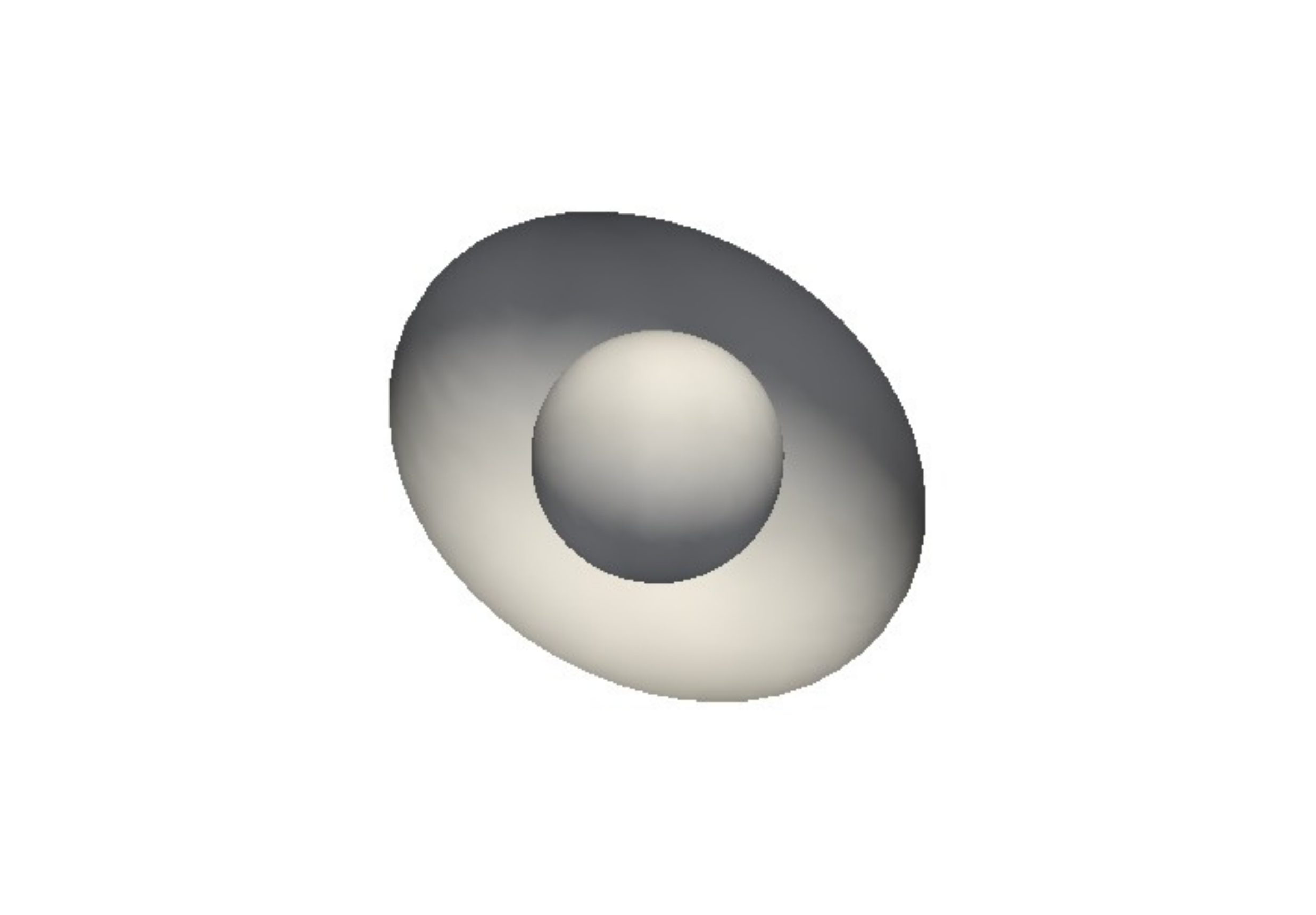}
    \label{sb31}
    \put(-60,0){\footnotesize $(a)$}
    \put(-120,33){\footnotesize $(iii)$}
}
    \subfigure{
    \includegraphics[scale=.148]{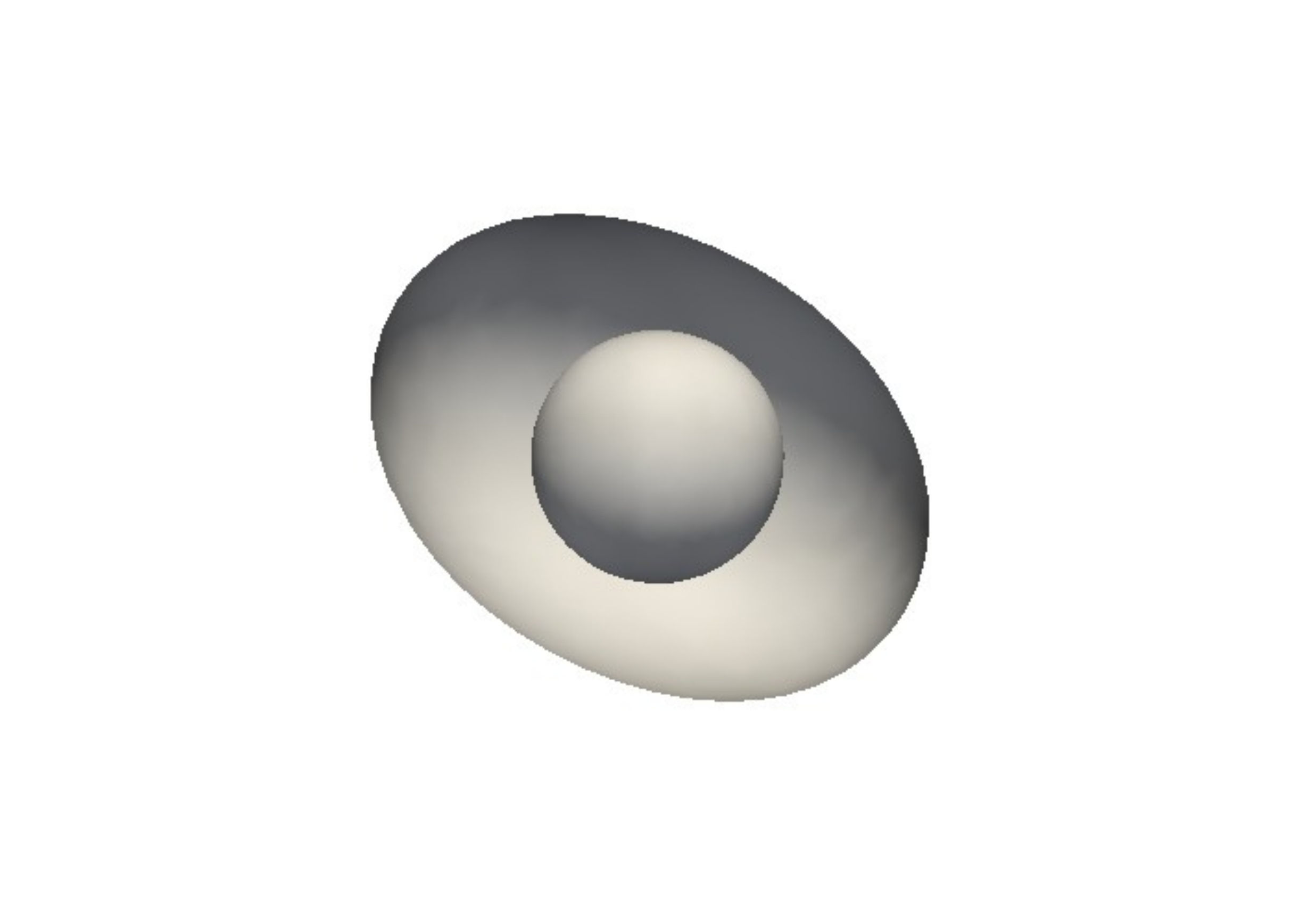}
    \label{sb32}
    \put(-60,0){\footnotesize $(b)$}
}
    \subfigure{
    \includegraphics[scale=.148]{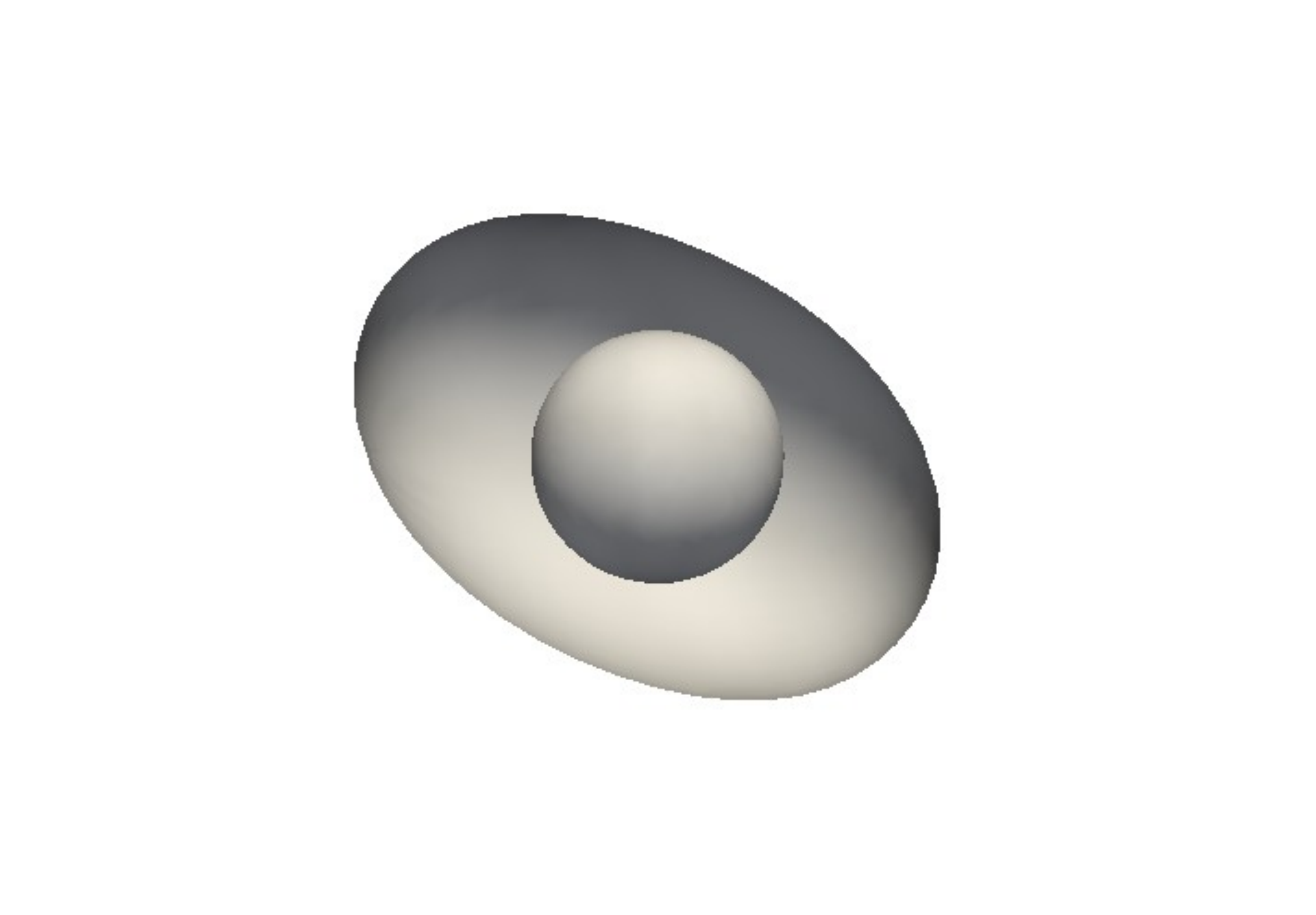}
    \label{sb33}
    \put(-60,0){\footnotesize $(c)$}
}
    \caption{Steady-state shape of an elastic capsule in shear flow at $Re=0.1$.  (i) cell without nucleus, (ii) with nucleus and bending stiffness $B=0$,  (iii) with nucleus and $B=10$. 
    The Capillary number is  (a) $Ca=0.15$,  (b) $Ca=0.30$ and  (c) $Ca=0.60$. }
        \label{f5}
\end{figure}

    The first two rows of figure \ref{f5}(a) depict the steady shape of the cell for three different capillary numbers of the outer membrane and zero beding stiffness. Note that, in the first row, the cell considered has no nucleus. In the absence of nucleus, the cell assumes an ellipsoidal shape, while it has a thicker middle part in the presence of the nucleus. Cells with nucleus thus have a lower flexibility and may encounter  more difficulties to pass through narrow vessels.

    The effect of bending stiffness on the deformation parameter is presented in figure \ref{f4}(b). It can be observed that cells with bending stiffness deform less and the reduction measured by the deformation parameter increases with the Capillary number. This effect is documented by the shape of capsules with bending stiffness reported in the lowest panels of figure \ref{f5}. Here, one can see that the deformation is reduced mainly on the edges of the capsule. Indeed, figure \ref{f5} shows that the cell shape is closer to an ellipsoid when adding bending rigidity. The difference between the different cases shown in the figures demonstrates that the effect of the bending rigidity is not negligible in such conditions and should be accounted for to obtain more accurate predictions.

\begin{figure}
    \centering
    \subfigure[]{\includegraphics[width=.5\columnwidth]{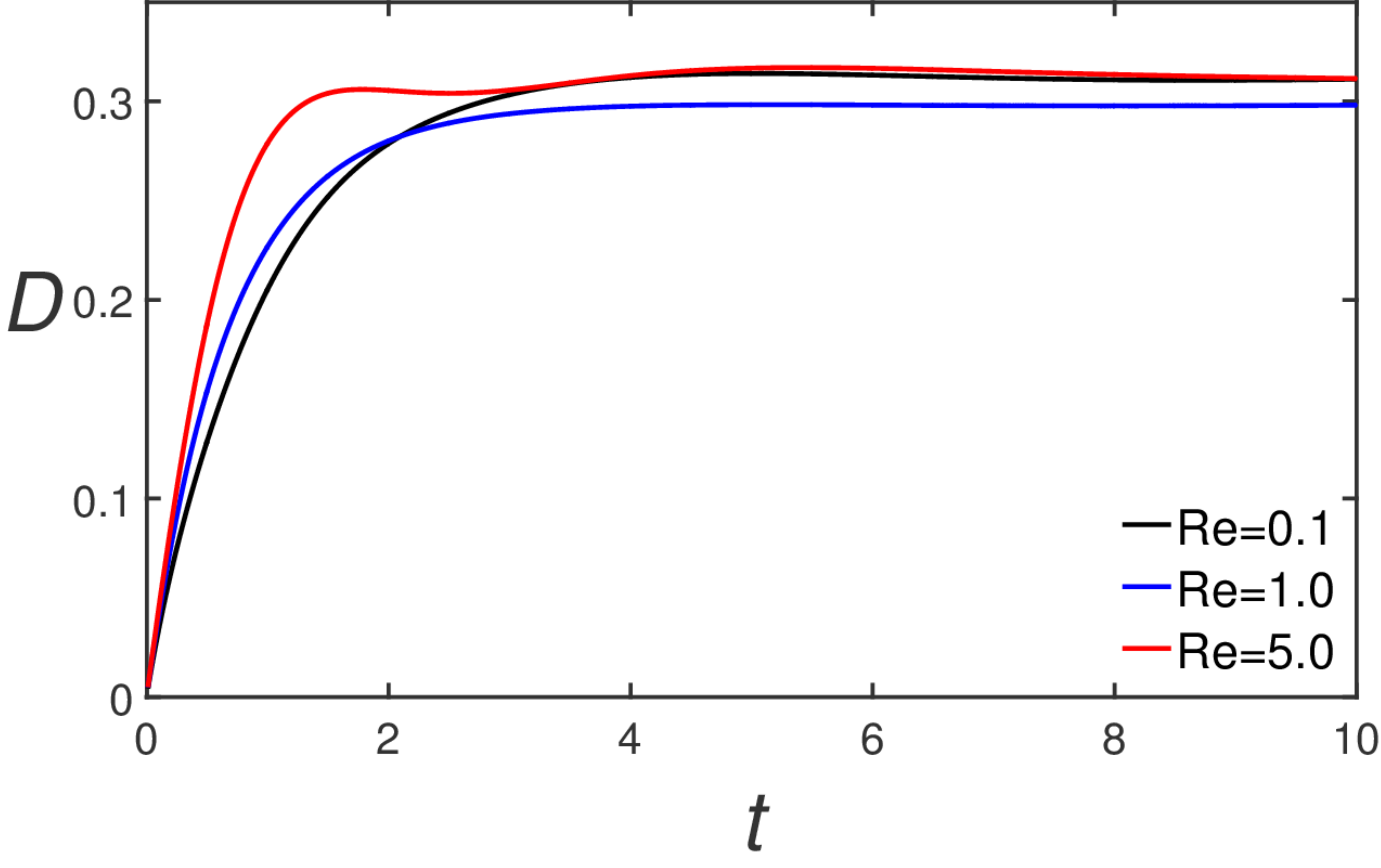}}\\
    \subfigure[]{\includegraphics[width=.35\columnwidth]{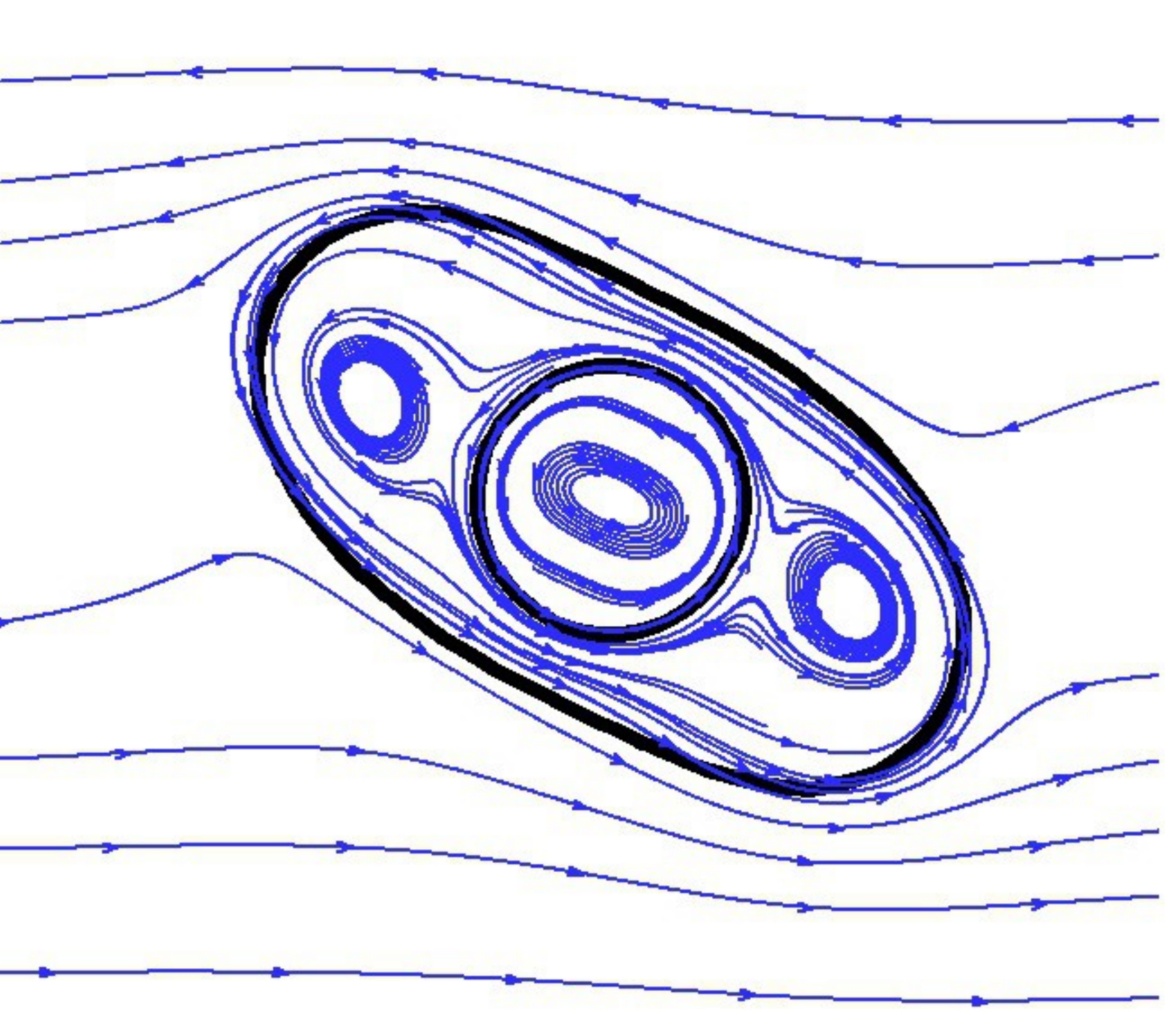}}
    \caption{ (a) Time evolution of deformation parameter for an initially spherical cell with nucleus for different Reynolds numbers, (b) streamlines in the mid $x_1$ plane for Re=5 at \textit{t}=2.5, black lines indicate location of the inner and outer membranes, $Ca=0.3$, $B=0.1$}
    \label{Re}
\end{figure}

    For a number of microfluidic applications, it may be important to understand the effect of flow inertia on the deformation of the transported cells. We therefore report in figure \ref{Re} the effect of increasing the Reynolds number on the deformation parameter. To prevent buckling, we considered small bending stiffness in the simulations. It can be observed that when increasing the Reynolds number the steady state deformation parameter first decreases for $Re=1$ and then increases ($Re=5$).
    The initial deformation rate is faster when increasing inertia. 
    Note also some oscillations in the deformation for $Re=5$, as observed in previous studies. These can be attributed to the formation of a pair of vortices inside the cell, on the two sides of the nucleus, in the transient stage (See figure 9b). Such vortices disappear at steady state but their formation and breakup results in oscillations of the cell membrane.   
 %   \textcolor{blue}{can we show the velocity field with the two vortices then?} \textcolor{green}{yes we just need to plot streamlines do it now or }

    \subsection{Rigid nucleus}
 
\begin{figure}
\centering
\includegraphics[width=.5\columnwidth]{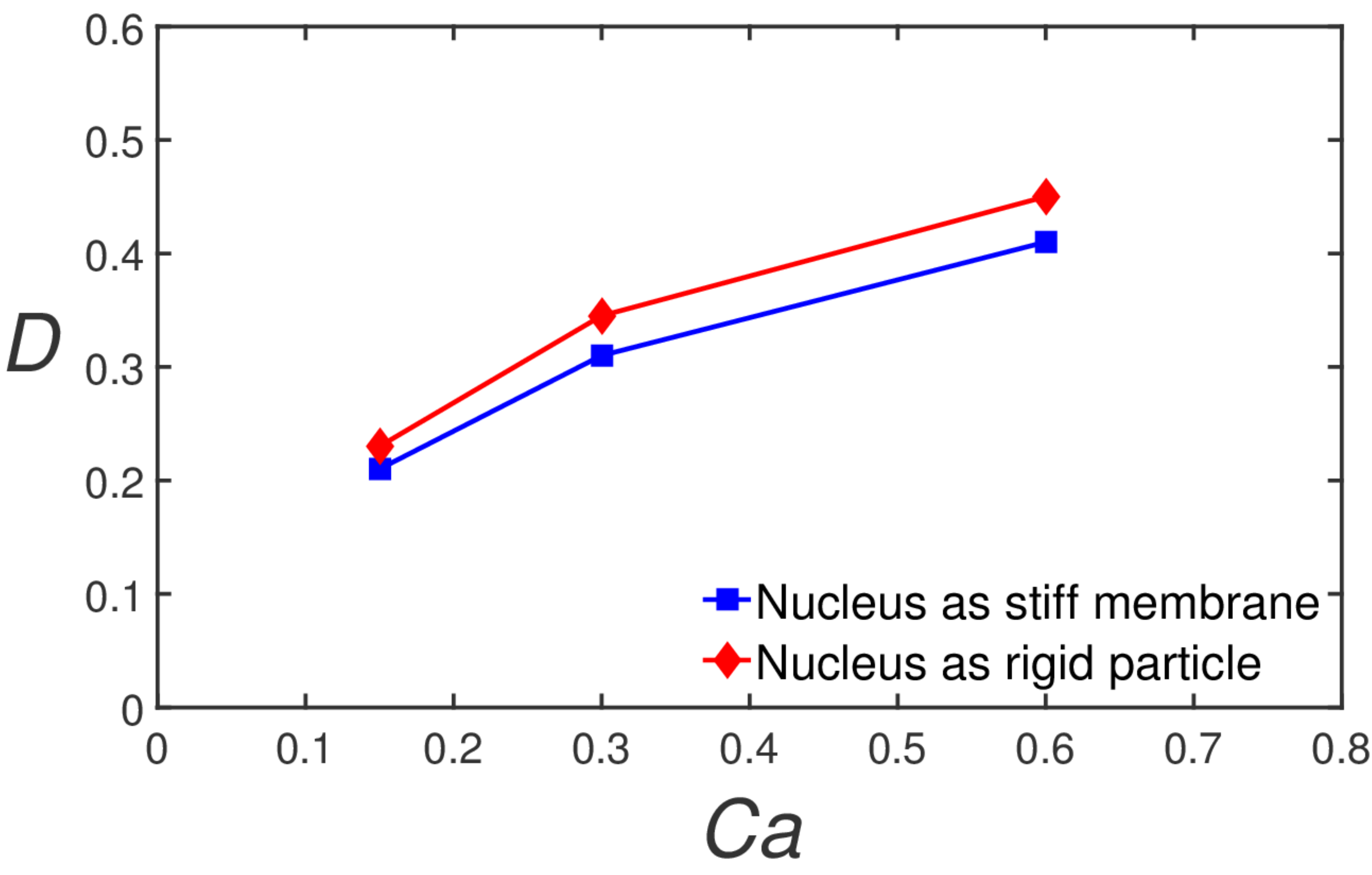} 
\caption{ Deformation parameter versus the Capillary number for capsules whose nucleus is modelled as a stiff membrane or as a rigid spherical particle, $\lambda=1$, $Re=0.1$. }
\label{f7}
\end{figure}

    As mentioned previously, two different models of nucleus are considered. For the results in the present section, the nucleus is modeled as a rigid, not deformable, particle following a different implementation of the immersed boundary method, see above. Figure \ref{f7} reports a comparison of the deformation for cells whose nucleus is modelled as a stiff membrane or as a rigid spherical particle, for viscosity ratio $\lambda=1$. The two methods produce approximately similar results, however with slightly larger deformation for cells with a rigid spherical nucleus. This fact can be related to the different nucleus rotation rates obtained with the two models. Indeed, for a rigid nucleus we assume no slip at the interface, whereas some slip is present at the surface of the nucleus if this is represented by an elastic membrane.
    
%    The main difference between the two approaches relies in the computational time. Indeed, 
    Finally, we report that the computational time needed for the case of a capsule with rigid nucleus is about 1.2 times that for a cell with a nucleus represented by a stiff membrane. Note that  the simulations assuming a rigid nucleus have been performed by coupling together two different approaches, able to model deformable and rigid object. This implementation opens possibility of modeling new more complicated structures, which will be investigated in the future.

\section{Conclusions}
\label{sec: conclusion}

The deformation of a capsule containing a stiff nucleus in homogeneous shear flow is studied numerically using an immersed boundary method to account for the fluid-solid interaction. The Neo-Hookean hyperelastic constitutive model is used to describe the cell membrane deformation while the fluid inside and outside each capsule is assumed to be Newtonian. The cell nucleus is modeled in two different ways, first as a second inner capsule with a significantly stiffer membrane and then as spherical rigid particle using a different implementation of the immersed boundary method, most suited to solid objects \cite{lashgari2016channel}.

In the immersed boundary method, a Lagrangian mesh is used to follow the deformation of the elastic membrane defining the capsule and an Eulerian mesh to solve the momentum equations. The shape of the membrane, its deformation and internal stresses are represented by means of spherical harmonics in order to have an accurate computation of the high-order derivatives of the membrane geometry. To save computational time in cases with very fine underlying Eulerian meshes, we have implemented the possibility of using a coarser Lagrangian mesh for the computation of the cell shape and stresses and a finer mesh for the communication of forces exchanged with the fluid. Spectral interpolation is employed to link the two Lagrangian representations of the geometry of the cell. Finally, to have the possibility to consider different viscosities inside and outside the cell, an indicator function is computed on the Eulerian mesh as the solution of a Poisson equation. The right hand side is obtained by spreading the normal vectors to cell surface known at the Lagrangian grid points onto the Eulerian mesh. The viscosity is then taken to be a function of this indicator function. The accuracy of the code is validated against results pertaining to the deformation of capsules without nucleus. In particular, we consider the inertialess flow of a capsule in homogeneous shear and  transport in pressure-driven Poiseuille flow at moderate Reynolds numbers (finite inertia).

The behavior of cells with a stiff nucleus in homogeneous shear has then been investigated. The cell deformation parameter is reported for different Capillary numbers, two values of the viscosity ratio and with or without bending rigidity. We observe that the deformation is smaller for cells with a nucleus. Examining the shape of the cell, that with nucleus is thicker in the middle part making it less flexible to pass through narrow vessels. When also considering bending stiffness, we observe an even smaller deformation and the shape of the cell is more regular and closer to an ellipsoid. Finally, we have compared the results obtained by modeling the nucleus as a rigid particle, reporting small differences. We show also that  the numerical approaches for rigid and deformable objects can coexist, which opens the possibility of modelling more complicated structures, e.g.\ small rigid cavities and obstacles in the flow \cite{zhu2014microfluidic}.

The method presented here can be employed and extended in the future to study the behavior of cells in different and more complicated configurations, enabling us to extract qualitative and quantitative data about the maximum stress on the membrane. Possible extensions include the possibility to consider a vesicle \cite{seol2016immersed}, i.e.\ an inextensible membrane, and a density contrast. Including cell-cell and cell-wall interactions in the numerical platform would allow us to study pair interactions of cells with nucleus in shear flow \cite{pranay2010pair} and ultimately investigate dense suspensions of deformable objects \cite{gao2011rheology,gao2013dynamics,kruger2014interplay}.

\section*{Acknowledgements}

This work was supported by the European Research Council Grant No. ERC-2013-CoG-616186, TRITOS and by the Swedish Research Council (VR). The authors acknowledge computer time provided by SNIC (Swedish National Infrastructure for Computing) and the support from the COST Action MP1305: Flowing matter. Dr.\ Lailai Zhu is thanked for the help with the implementation of the membrane module.

%\bibliography{references}

\begin{thebibliography}{10}

\bibitem{adams1999spherepack}
John~C Adams and Paul~N Swarztrauber.
\newblock Spherepack 3.0: A model development facility.
\newblock {\em Monthly Weather Review}, 127(8):1872--1878, 1999.

\bibitem{ardekani2016numerical}
Mehdi~Niazi Ardekani, Pedro Costa, Wim-Paul Breugem, and Luca Brandt.
\newblock Numerical study of the sedimentation of spheroidal particles.
\newblock {\em arXiv preprint arXiv:1602.05769}, 2016.

\bibitem{bannister2003ins}
Lawrence Bannister and Graham Mitchell.
\newblock The ins, outs and roundabouts of malaria.
\newblock {\em Trends in parasitology}, 19(5):209--213, 2003.

\bibitem{breugem2012second}
Wim-Paul Breugem.
\newblock A second-order accurate immersed boundary method for fully resolved
  simulations of particle-laden flows.
\newblock {\em Journal of Computational Physics}, 231(13):4469--4498, 2012.

\bibitem{caille1998assessment}
Nathalie Caille, Yanik Tardy, and Jean-Jacques Meister.
\newblock Assessment of strain field in endothelial cells subjected to uniaxial
  deformation of their substrate.
\newblock {\em Annals of biomedical engineering}, 26(3):409--416, 1998.

\bibitem{caille2002contribution}
Nathalie Caille, Olivier Thoumine, Yanik Tardy, and Jean-Jacques Meister.
\newblock Contribution of the nucleus to the mechanical properties of
  endothelial cells.
\newblock {\em Journal of biomechanics}, 35(2):177--187, 2002.

\bibitem{chang1993experimental}
Kuo-Shu Chang and William~Lee Olbricht.
\newblock Experimental studies of the deformation and breakup of a synthetic
  capsule in steady and unsteady simple shear flow.
\newblock {\em Journal of Fluid Mechanics}, 250:609--633, 1993.

\bibitem{chorin1968numerical}
Alexandre~Joel Chorin.
\newblock Numerical solution of the navier-stokes equations.
\newblock {\em Mathematics of computation}, 22(104):745--762, 1968.

\bibitem{cooke2001malaria}
Brian~M Cooke, Narla Mohandas, and Ross~L Coppel.
\newblock The malaria-infected red blood cell: structural and functional
  changes.
\newblock {\em Advances in parasitology}, 50:1--86, 2001.

\bibitem{dodd2014fast}
Michael~S Dodd and Antonino Ferrante.
\newblock A fast pressure-correction method for incompressible two-fluid flows.
\newblock {\em Journal of Computational Physics}, 273:416--434, 2014.

\bibitem{fischer1977tank}
Th~Fischer.
\newblock Tank tread motion of red-cell membranes in viscometric flow-behavior
  of intracellular and extracellular markers (with film).
\newblock {\em Blood cells}, 3(2):351--365, 1977.

\bibitem{fischer1978red}
Thomas~M Fischer, M~Stohr-Lissen, and Holger Schmid-Schonbein.
\newblock The red cell as a fluid droplet: tank tread-like motion of the human
  erythrocyte membrane in shear flow.
\newblock {\em Science}, 202(4370):894--896, 1978.

\bibitem{freund2010high}
JB~Freund and H~Zhao.
\newblock A high-resolution fast boundary-integral method for multiple
  interacting blood cells.
\newblock {\em Computational Hydrodynamics of Capsules and Biological Cells},
  page~71, 2010.

\bibitem{freund2007leukocyte}
Jonathan~B Freund.
\newblock Leukocyte margination in a model microvessel.
\newblock {\em Physics of Fluids (1994-present)}, 19(2):023301, 2007.

\bibitem{gaehtgens1979motion}
P~Gaehtgens, C~D{\"u}hrssen, and KH~Albrecht.
\newblock Motion, deformation, and interaction of blood cells and plasma during
  flow through narrow capillary tubes.
\newblock {\em Blood cells}, 6(4):799--817, 1979.

\bibitem{galbraith1998shear}
CG~Galbraith, R~Skalak, and S~Chien.
\newblock Shear stress induces spatial reorganization of the endothelial cell
  cytoskeleton.
\newblock {\em Cell motility and the cytoskeleton}, 40(4):317--330, 1998.

\bibitem{gao2011rheology}
Tong Gao, Howard~H Hu, and Pedro~Ponte Casta{\~n}eda.
\newblock Rheology of a suspension of elastic particles in a viscous shear
  flow.
\newblock {\em Journal of Fluid Mechanics}, 687:209, 2011.

\bibitem{gao2013dynamics}
Tong Gao, Howard~H Hu, and Pedro~Ponte Casta{\~n}eda.
\newblock Dynamics and rheology of elastic particles in an extensional flow.
\newblock {\em Journal of Fluid Mechanics}, 715:573--596, 2013.

\bibitem{goldsmith1972flow}
HL~Goldsmith and Jean Marlow.
\newblock Flow behaviour of erythrocytes. i. rotation and deformation in dilute
  suspensions.
\newblock {\em Proceedings of the Royal Society of London B: Biological
  Sciences}, 182(1068):351--384, 1972.

\bibitem{guilak1995compression}
Farshid Guilak.
\newblock Compression-induced changes in the shape and volume of the
  chondrocyte nucleus.
\newblock {\em Journal of biomechanics}, 28(12):1529--1541, 1995.

\bibitem{guilak2000mechanical}
Farshid Guilak and Van~C Mow.
\newblock The mechanical environment of the chondrocyte: a biphasic finite
  element model of cell--matrix interactions in articular cartilage.
\newblock {\em Journal of biomechanics}, 33(12):1663--1673, 2000.

\bibitem{guo2016deformability}
Quan Guo, Simon~P Duffy, Kerryn Matthews, Xiaoyan Deng, Aline~T Santoso, Emel
  Islamzada, and Hongshen Ma.
\newblock Deformability based sorting of red blood cells improves diagnostic
  sensitivity for malaria caused by plasmodium falciparum.
\newblock {\em Lab on a Chip}, 16(4):645--654, 2016.

\bibitem{huang2012three}
Wei-Xi Huang, Cheong~Bong Chang, and Hyung~Jin Sung.
\newblock Three-dimensional simulation of elastic capsules in shear flow by the
  penalty immersed boundary method.
\newblock {\em Journal of Computational Physics}, 231(8):3340--3364, 2012.

\bibitem{ingber1990fibronectin}
Donald~E Ingber.
\newblock Fibronectin controls capillary endothelial cell growth by modulating
  cell shape.
\newblock {\em Proceedings of the National Academy of Sciences},
  87(9):3579--3583, 1990.

\bibitem{kan1999effects}
Heng-Chuan Kan, Wei Shyy, HS~Udaykumar, Philippe Vigneron, and Roger
  Tran-Son-Tay.
\newblock Effects of nucleus on leukocyte recovery.
\newblock {\em Annals of biomedical engineering}, 27(5):648--655, 1999.

\bibitem{kessler2008swinging}
S~Kessler, R~Finken, and U~Seifert.
\newblock Swinging and tumbling of elastic capsules in shear flow.
\newblock {\em Journal of Fluid Mechanics}, 605:207--226, 2008.

\bibitem{kilimnik2011inertial}
Alex Kilimnik, Wenbin Mao, and Alexander Alexeev.
\newblock Inertial migration of deformable capsules in channel flow.
\newblock {\em Physics of Fluids (1994-present)}, 23(12):123302, 2011.

\bibitem{kim2015inertial}
Boyoung Kim, Cheong~Bong Chang, Sung~Goon Park, and Hyung~Jin Sung.
\newblock Inertial migration of a 3d elastic capsule in a plane poiseuille
  flow.
\newblock {\em International Journal of Heat and Fluid Flow}, 54:87--96, 2015.

\bibitem{kruger2014interplay}
Timm Kr{\"u}ger, Badr Kaoui, and Jens Harting.
\newblock Interplay of inertia and deformability on rheological properties of a
  suspension of capsules.
\newblock {\em Journal of Fluid Mechanics}, 751:725--745, 2014.

\bibitem{lac2005deformation}
Etienne Lac and Dominique Barth{\`e}s-Biesel.
\newblock Deformation of a capsule in simple shear flow: effect of membrane
  prestress.
\newblock {\em Physics of Fluids (1994-present)}, 17(7):072105, 2005.

\bibitem{lashgari2014laminar}
Iman Lashgari, Francesco Picano, Wim-Paul Breugem, and Luca Brandt.
\newblock Laminar, turbulent, and inertial shear-thickening regimes in channel
  flow of neutrally buoyant particle suspensions.
\newblock {\em Physical review letters}, 113(25):254502, 2014.

\bibitem{lashgari2016channel}
Iman Lashgari, Francesco Picano, Wim~Paul Breugem, and Luca Brandt.
\newblock Channel flow of rigid sphere suspensions: Particle dynamics in the
  inertial regime.
\newblock {\em International Journal of Multiphase Flow}, 78:12--24, 2016.

\bibitem{li20102decomp}
Ning Li and Sylvain Laizet.
\newblock 2decomp \& fft-a highly scalable 2d decomposition library and fft
  interface.
\newblock In {\em Cray User Group 2010 conference}, pages 1--13, 2010.

\bibitem{li2008front}
Xiaoyi Li and Kausik Sarkar.
\newblock Front tracking simulation of deformation and buckling instability of
  a liquid capsule enclosed by an elastic membrane.
\newblock {\em Journal of Computational Physics}, 227(10):4998--5018, 2008.

\bibitem{lim2006mechanical}
CT~Lim, EH~Zhou, and ST~Quek.
\newblock Mechanical models for living cells--a review.
\newblock {\em Journal of biomechanics}, 39(2):195--216, 2006.

\bibitem{maniotis1997demonstration}
Andrew~J Maniotis, Christopher~S Chen, and Donald~E Ingber.
\newblock Demonstration of mechanical connections between integrins,
  cytoskeletal filaments, and nucleoplasm that stabilize nuclear structure.
\newblock {\em Proceedings of the National Academy of Sciences},
  94(3):849--854, 1997.

\bibitem{peskin2002immersed}
Charles~S Peskin.
\newblock The immersed boundary method.
\newblock {\em Acta numerica}, 11:479--517, 2002.

\bibitem{pozrikidis1995finite}
C~Pozrikidis.
\newblock Finite deformation of liquid capsules enclosed by elastic membranes
  in simple shear flow.
\newblock {\em Journal of Fluid Mechanics}, 297:123--152, 1995.

\bibitem{pozrikidis2001effect}
C~Pozrikidis.
\newblock Effect of membrane bending stiffness on the deformation of capsules
  in simple shear flow.
\newblock {\em Journal of Fluid Mechanics}, 440:269--291, 2001.

\bibitem{pozrikidis2010computational}
Constantine Pozrikidis.
\newblock {\em Computational hydrodynamics of capsules and biological cells}.
\newblock CRC Press, 2010.
\newblock p. 89.

\bibitem{pranay2010pair}
Pratik Pranay, Samartha~G Anekal, Juan~P Hernandez-Ortiz, and Michael~D Graham.
\newblock Pair collisions of fluid-filled elastic capsules in shear flow:
  Effects of membrane properties and polymer additives.
\newblock {\em Physics of Fluids (1994-present)}, 22(12):123103, 2010.

\bibitem{ramanujan1998deformation}
S~Ramanujan and C~Pozrikidis.
\newblock Deformation of liquid capsules enclosed by elastic membranes in
  simple shear flow: large deformations and the effect of fluid viscosities.
\newblock {\em Journal of Fluid Mechanics}, 361:117--143, 1998.

\bibitem{rodriguez2013review}
Marita~L Rodriguez, Patrick~J McGarry, and Nathan~J Sniadecki.
\newblock Review on cell mechanics: experimental and modeling approaches.
\newblock {\em Applied Mechanics Reviews}, 65(6):060801, 2013.

\bibitem{roma1999adaptive}
Alexandre~M Roma, Charles~S Peskin, and Marsha~J Berger.
\newblock An adaptive version of the immersed boundary method.
\newblock {\em Journal of computational physics}, 153(2):509--534, 1999.

\bibitem{rorai2015motion}
Cecilia Rorai, Antoine Touchard, Lailai Zhu, and Luca Brandt.
\newblock Motion of an elastic capsule in a constricted microchannel.
\newblock {\em The European Physical Journal E}, 38(5):1--13, 2015.

\bibitem{schmid1969fluid}
Holger Schmid-Sch{\"o}nbein and Roe Wells.
\newblock Fluid drop-like transition of erythrocytes under shear.
\newblock {\em Science}, 165(3890):288--291, 1969.

\bibitem{seol2016immersed}
Yunchang Seol, Wei-Fan Hu, Yongsam Kim, and Ming-Chih Lai.
\newblock An immersed boundary method for simulating vesicle dynamics in three
  dimensions.
\newblock {\em Journal of Computational Physics}, 322:125--141, 2016.

\bibitem{skalak1969deformation}
R~Skalak and PI~Branemark.
\newblock Deformation of red blood cells in capillaries.
\newblock {\em Science}, 164(3880):717--719, 1969.

\bibitem{skotheim2007red}
JM~Skotheim and Timothy~W Secomb.
\newblock Red blood cells and other nonspherical capsules in shear flow:
  oscillatory dynamics and the tank-treading-to-tumbling transition.
\newblock {\em Physical review letters}, 98(7):078301, 2007.

\bibitem{swarztrauber2000generalized}
Paul~N Swarztrauber and William~F Spotz.
\newblock Generalized discrete spherical harmonic transforms.
\newblock {\em Journal of Computational Physics}, 159(2):213--230, 2000.

\bibitem{uhlmann2005immersed}
Markus Uhlmann.
\newblock An immersed boundary method with direct forcing for the simulation of
  particulate flows.
\newblock {\em Journal of Computational Physics}, 209(2):448--476, 2005.

\bibitem{unverdi1992front}
Salih~Ozen Unverdi and Gr{\'e}tar Tryggvason.
\newblock A front-tracking method for viscous, incompressible, multi-fluid
  flows.
\newblock {\em Journal of computational physics}, 100(1):25--37, 1992.

\bibitem{walter2001shear}
Anja Walter, Heinz Rehage, and Herbert Leonhard.
\newblock Shear induced deformation of microcapsules: shape oscillations and
  membrane folding.
\newblock {\em Colloids and Surfaces A: Physicochemical and Engineering
  Aspects}, 183:123--132, 2001.

\bibitem{walter2010coupling}
J~Walter, A-V Salsac, D~Barth{\`e}s-Biesel, and P~Le~Tallec.
\newblock Coupling of finite element and boundary integral methods for a
  capsule in a stokes flow.
\newblock {\em International journal for numerical methods in engineering},
  83(7):829--850, 2010.

\bibitem{wu2013simulation}
Tenghu Wu and James~J Feng.
\newblock Simulation of malaria-infected red blood cells in microfluidic
  channels: Passage and blockage.
\newblock {\em Biomicrofluidics}, 7(4):044115, 2013.

\bibitem{zhang2015multiple}
Yao Zhang, Changjin Huang, Sangtae Kim, Mahdi Golkaram, Matthew~WA Dixon, Leann
  Tilley, Ju~Li, Sulin Zhang, and Subra Suresh.
\newblock Multiple stiffening effects of nanoscale knobs on human red blood
  cells infected with plasmodium falciparum malaria parasite.
\newblock {\em Proceedings of the National Academy of Sciences},
  112(19):6068--6073, 2015.

\bibitem{zhao2010spectral}
Hong Zhao, Amir~HG Isfahani, Luke~N Olson, and Jonathan~B Freund.
\newblock A spectral boundary integral method for flowing blood cells.
\newblock {\em Journal of Computational Physics}, 229(10):3726--3744, 2010.

\bibitem{zhu2015motion}
Lailai Zhu and Luca Brandt.
\newblock The motion of a deforming capsule through a corner.
\newblock {\em Journal of Fluid Mechanics}, 770:374--397, 2015.

\bibitem{zhu2015dynamics}
LaiLai Zhu, Jean Rabault, and Luca Brandt.
\newblock The dynamics of a capsule in a wall-bounded oscillating shear flow.
\newblock {\em Physics of Fluids (1994-present)}, 27(7):071902, 2015.

\bibitem{zhu2014microfluidic}
Lailai Zhu, Cecilia Rorai, Dhrubaditya Mitra, and Luca Brandt.
\newblock A microfluidic device to sort capsules by deformability: a numerical
  study.
\newblock {\em Soft matter}, 10(39):7705--7711, 2014.

\end{thebibliography}
\bibliographystyle{plain}

\end{document}